\documentclass[
reprint, 
 amsmath,amssymb,
 aps, 
]{revtex4-2}

\usepackage{graphicx}
\usepackage{dcolumn}
\usepackage{bm}

\usepackage{siunitx}
\DeclareSIUnit\bw{0.1 \%~bandwidth}
\DeclareSIUnit\columb{C}
\DeclareSIUnit\photon{ph}

\usepackage{float}
\usepackage{enumitem}
\usepackage{layouts}
\usepackage{placeins}

\begin{document}


\title{\textbf{Characterizing an inverse Compton X-ray source and determining its electron beam parameters using a genetic algorithm} 
}%

\author{Johannes Melcher}
 \altaffiliation{contributed equally}
\author{Jen-Fu Tu}%
\altaffiliation{contributed equally}
\author{Franz Pfeiffer}
\altaffiliation[Also at: ]{Institute for Diagnostic and Interventional Radiology, TUM School of Medicine and Health, TUM Klinikum, Technical University of Munich, 81675 Munich, Germany \& TUM Institute for Advanced Study, Technical University of Munich, 85748 Garching, Germany}
\author{Martin Dierolf}
\author{Benedikt G\"unther}
 \email{Contact author: benedikt.guenther@tum.de}
\affiliation{%
 Chair of Biomedical Physics, TUM School of Natural Sciences, Technical University of Munich, 85748 Garching, Germany}
\affiliation{Munich Institute of Biomedical Engineering, Technical University of Munich, 85748 Garching, Germany}

\author{Bal{\v s}a Terzi\'c}
\author{Erik Johnson}

\affiliation{
Center for Accelerator Science, Department of Physics, Old Dominion University, Norfolk, VA 23529, USA
}%

\author{Geoffrey Krafft}
\affiliation{%
 Thomas Jefferson National Accelerator Facility, Newport News, VA 23606, USA
}%
\affiliation{
Center for Accelerator Science, Department of Physics, Old Dominion University, Norfolk, VA 23529, USA
}%
\date{\today}

\begin{abstract}
Inverse Compton X-ray sources are laboratory-scale devices providing quasi-monochromatic synchrotron radiation which is generated by laser photons Compton-scattering off highly relativistic electrons.
Since the shape and width of the X-ray spectrum are determined by the properties of the colliding beams, these must be carefully optimised.
However, device compactness limits the space for diagnostics, rendering a complete characterisation challenging, especially if an electron storage ring is combined with a laser enhancement cavity.
Here, a framework for laser, electron and X-ray beam parameter determination is proposed to address this issue.
First, methods for determining the laser- and X-ray parameters are presented.
Knowing these, electron beam parameters are retrieved from the shape of the X-ray spectrum.
To this end, an analytical physical model enabling a rapid calculation of inverse Compton scattering spectra is developed and combined with a genetic algorithm.
This strategy’s effectiveness is demonstrated by applying the concept at the Munich Compact Light Source, a storage ring-based inverse Compton X-ray source facility.
Since the analytical model is computationally very inexpensive, the proposed framework could enable real-time monitoring of inverse Compton X-ray sources or be used as a non-invasive diagnostic based on a single spectrum for the electron beam emittance of storage rings or accelerators.
\end{abstract}

\maketitle


\section{Introduction}
Brilliant, i.e.\ highly-monochromatic \& partially-coherent, X-rays are a powerful diagnostic probe.
They can be employed for high-resolution (phase-contrast) X-ray computed tomography and microscopy, e.g.\ \cite[Chapter 13]{Gunther2023}, wide- and small-angle scattering, e.g.\ \cite{Melcher2024}, or spectroscopy, e.g.\ \cite[Chapter 13]{Gunther2023}, to name only a few techniques.
Most of these techniques were developed at and still mainly are used at large-scale synchrotron facilities because the lower brilliance of X-ray tube sources limits their transfer into local laboratories.
This limitation, in conjunction with the emergence of high-power commercial laser systems, has sparked the research on so-called inverse Compton X-ray sources (ICS) in the last decade which bridge the gap in brilliance between synchrotrons and laboratory X-ray tube sources.
Accordingly, many inverse Compton X-ray sources are currently under development, in commissioning, or already operational \cite[Chapter 6]{Gunther2023}.
Among them are the Arizona State University's Compact X-ray Light Source (CXLS)\cite{Graves2023}, BriXsino \cite{Drebot2019}, STAR \cite{Bacci2014,Bacci2016}, ThomX \cite{Jacquet2024}, Tsinghua Thompson X-ray Source \cite{Chi2018}, the ICS Project at UC Irvine / Lumitron \cite{Barty2024}, Smart*Light \cite{Luiten2016} or the Munich Compact Light Source (MuCLS)\cite{Gunther2023}.
The latter has been one of the very first ICS-based user facilities and the experiments presented in this study were performed there.\\
The collision of a highly-relativistic electron beam (typically on the order of tens of MeV) with a laser beam produces quasi-monochromatic X-ray photons in ICSs.
The quality of both, the electron and the laser beam, significantly affect the brilliance of the generated X-ray beam in inverse Compton scattering.
Knowing the laser parameters, this causality can be exploited to retrieve the electron beam parameters by analysing the measured spectrum of the ICS's scattered radiation.
In the following, a framework for the characterisation of the MuCLS's laser, electron and X-ray beam parameters is presented.
First, a model for the inverse Compton scattering process is introduced allowing for rapid calculation of the emitted X-ray spectrum.
Afterwards, the strategy for the characterisation of an ICS is presented.
Following this, the MuCLS is described and the laser and X-ray parameters are retrieved.
In the last step, these parameters and the mathematical model are used to retrieve the electron beam parameters with a genetic algorithm that tries to optimally model the measured X-ray spectrum by varying electron beam emittance.
It allows for the emittance to be different in the two transverse degrees of freedom, which is an important condition \cite{HAJIMA2021}.\\
While this procedure is applicable to any accelerator, ICSs may benefit most because their compact footprint results in scarce space for diagnostic devices, especially in storage ring-based ICSs.
This strongly limits the characterisation of their interacting electron and laser beams.
Lastly, this technique could be used as a diagnostic tool for non-invasive permanent emittance monitoring in ICSs or storage rings.

\section{A quantitative model for calculating inverse Compton X-ray spectra transmitted by a finite aperture}\label{sec:theory}

The differential cross section for (inverse) Compton scattering in the rest frame of the electron is governed by the Klein-Nishina formula, which accounts for electron recoil \cite{Jackson1975,Berestetskii1980}.
A number of numerical codes, each with its own approximations, have been developed by various groups to compute spectra emerging from ICSs in both the low laser-field regime (when the amplitude of the normalized vector potential of the laser pulse $a_0 \ll 1$), e.g., \cite{Sun2011,Petrillo2012,krafft16, ranjan18}, and the high laser-field regime ($a_0 \gtrapprox 1$), e.g., \cite{Martens2021,CAIN,Tomassini2005,Krafft2023}.\\
The radiation spectra emerging from ICSs share common features: the high-energy side of the spectrum (Compton edge) is sharp, with its Gaussian broadening directly related to the electron beam energy spread; 
and the low-energy side of the spectrum falls off exponentially, with the slope determined by the electron beam's emittance and the beam size.
These common features strongly suggest that a simplification in the spectral calculation is possible.
Here we outline such a simplified mathematical formalism for efficient computation of radiation spectra emerging from ICSs which exploits the following approximations:
\begin{enumerate}[nosep, label=(\roman*), ref=\roman*]
    \item Gaussian spatial and energy distribution of the laser beam;
    \item Gaussian phase space distribution of the electron beam; 
    \item $\alpha_x = \alpha_y = 0$, i.e.\ emittance ellipse aligned with coordinate system; \label{item:alpha_eq_zero}
    \item a head-on collision of laser and electron beam;\label{item:head_on}
    \item an aperture centred on axis. Although this is not a restriction of the model, we chose this case because of the measurements. \label{item:aperture_on_axis}
\end{enumerate}
Invoking these approximations allows for analytic integration over electron beam emittances and a simple integration over electron beam energy spread. 
These simplifications effectively reduce the required computation of the radiation spectra to a 3D integration (2D aperture and 1D scattering angle), obviating cumbersome and computationally expensive multidimensional Monte Carlo integration over 3D electron beam phase-space parameters.
The resulting computer code is substantially more efficient than our previous, more general codes \cite{krafft16,ranjan18} to the point where the science reach of the new code extends our previous capabilities.
For example, and most pertinently to this work, integration of the new code into an iterative genetic algorithm, requiring many generations of individual simulations for convergence (see Section \ref{Sec_BeamParams}), takes only a couple of hours of computer time instead of weeks.
In this section, we report on the most salient results of the mathematical 
formalism whose derivation will be detailed in the upcoming publication 
\cite{Krafft2026}.

\subsection{An aperture-dependent reduction factor for Compton scattering}

For an individual electron and given finite aperture, a limited energy range exists over which the scattered photon spectrum can extend.
The energy range may be found through careful analysis of the Compton emission formula.
In general, in the Thomson limit \cite{Arutyunian1963}
	\[{E}'={{E}_{p}}\frac{1-\vec{\beta }\cdot \vec{k}}{1-\vec{\beta }\cdot {\vec{k}}'},\]
where ${{E}_{p}}$ is the incident laser photon (mean) energy, ${E}'$ is the scattered photon energy, $\vec{\beta }$ is the lab-frame velocity vector for the electron normalised by the velocity of light, and $\vec{k}$ and ${\vec{k}}'$ are the unit vectors along the propagation direction for the incident and scattered photons, respectively.
Assume the polar coordinates of the electron velocity are $\vec{\beta }=\beta \left( \sin {\theta }\cos {\phi },\sin {\theta }\sin {\phi },\cos {\theta } \right)$ and the scattered radiation has the usual polar coordinate description ${\vec{k}}'=\left( \sin \theta' \cos \phi' ,\sin \theta' \sin \phi' ,\cos \theta'  \right)$.
When the incident wave vector is anti-aligned with the $z$-axis, the relativistically correct Thomson energy formula evaluates to
\[{E}'={{E}_{p}}\frac{1+\beta \cos {\theta }}{1-\beta \left( \sin {\theta }'\sin \theta \cos \left( {\phi }'-\phi  \right)+\cos {\theta }'\cos \theta  \right)}.\]
Because $\beta <1$, the energy formula evaluates to a finite number for all choices of angle.
An important energy for quantifying the emission is 
$E_{\gamma,{\rm max}}=\left(1+\beta\right)^2 \gamma^2 E_p$, known as the Compton edge energy, which is the maximum energy emitted in Thomson scattering \cite{Arutyunian1963}.
Since photons of this energy are always emitted by every incident electron exactly in the forward direction, the only electrons that will add photons to the Compton edge in the small source limit are those whose angles fall within the angular range of any collecting aperture.
More generally, for a large source and an arbitrary location $(x_a,y_a)$ in the aperture plane, only those electrons with source location $(x_s,y_s)$ and angle $(\theta_x,\theta_y)$ satisfying the straight-line propagation conditions
	\[x_a=x_s+\theta_x L,\]
	\[y_a=y_s+\theta_y L,\]
will add to the Compton edge if radiation is collected at $(x_a,y_a)$, where $L$ is the distance between the interaction point and the aperture.

For Gaussian beams, these considerations lead to a straightforward estimate of the fraction of scattered photons contributing to the Compton edge value at each location in the aperture: 
If the normalized transverse phase space distribution of the electrons at the interaction point is
\begin{widetext}
	\[f_e(x_s,y_s,\theta_x,\theta_y)=\frac{\exp(-x_s^2/2\sigma_{e,x}^2)\exp(-y_s^2/2\sigma_{e,y}^2)\exp(-\theta_x^2/2\sigma_{\theta_x}^2)
	\exp(-\theta_y^2/2\sigma_{\theta_y}^2)}{(2\pi)^2\sigma_{e,x}\sigma_{e,y}\sigma_{\theta_x}\sigma_{\theta_y}},\]
\end{widetext}
and the normalized photon transverse intensity distribution is
	\[f_p(x_s,y_s)=\frac{\exp(-x_s^2/2\sigma_{p,x}^2)\exp(-y_s^2/2\sigma_{p,y}^2)}{2\pi\sigma_{p,x}\sigma_{p,y}},\]
the unit-normalized distribution of the photons at the Compton edge ($f_{CE}$), proportional to $f_ef_p$, is
\begin{widetext}
	\[f_{CE}(x_s,y_s,\theta_x,\theta_y)=\frac{\exp(-x_s^2/2\sigma^2_{s,x})\exp(-y_s^2/2\sigma^2_{s,y})}{2\pi\sigma_{s,x}\sigma_{s,y}}
	\frac{\exp(-\theta_x^2/2\sigma_{\theta_x}^2)
	\exp(-\theta_y^2/2\sigma_{\theta_y}^2)}{2\pi\sigma_{\theta_x}\sigma_{\theta_y}},\]
\end{widetext}
where the source {\it {rms}}-size in both directions $i=\{x,y\}$ is
\begin{equation}\label{eq:sigma_source_beams}
\sigma_{s,i}=\frac{\sigma_{e,i}\sigma_{p,i}}{\sqrt
{\sigma_{e,i}^2+\sigma_{p,i}^2}}.
\end{equation}
$\sigma_{p,i}$, $\sigma_{e,i}$ and $\sigma_{\theta_i}$ denote the transverse rms-size of the photon, electron beam and the latter's rms-angular distribution.

Using the propagation relations, the fraction of Compton-edge photons in a small area element in the aperture plane is
\begin{widetext}
	\[f_{CE}(x_a,y_a)dx_ady_a=\int \int \frac{\exp(-x_s^2/2\sigma^2_{s,x})\exp(-y_s^2/2\sigma^2_{s,y})}{2\pi\sigma_{s,x}\sigma_{s,y}}\]
	\[\frac{\exp(-(x_a-x_s)^2/2\sigma_{\theta_x}^2L^2)
	\exp(-(y_a-y_s)^2/2\sigma_{\theta_y}^2L^2)}{2\pi
	\sigma_{\theta_x}\sigma_{\theta_y}L^2}dx_sdy_sdx_ady_a.\]
\end{widetext}
Integrating over the Gaussian source distributions by completing the squares yields
	\[f_{CE}(x_a,y_a)= \frac{\exp(-x_a^2/2\sigma^2_{f,x})\exp(-y_a^2/2\sigma^2_{f,y})}{2\pi\sigma_{f,x}\sigma_{f,y}},\]
where
\begin{equation} \label{eq:sigma_source_tot}
\sigma^2_{f,i}=\sigma^2_{s,i}+\sigma^2_{\theta_i}L^2.
\end{equation}
Consequently, an increase in the electron beam divergence reduces the flux into a given aperture.\\
Because the edge photon distribution was unit normalized, integrating $f_{\mathrm{CE}}$ over the whole aperture plane must yield 1.
Consequently, integrating over a finite aperture returns the fraction of Compton-edge photons transmitted through it. 
This fact will be used in a moment to obtain the overall scale of the spectral energy distribution of the scattered photons.

Consistent with our 2-dimensional modelling of the incident photon pulse, the spectral density at the Compton edge transmitted through the aperture can be defined as
\begin{equation}
{\left. {\frac{{dN}}{{d{E'}}}}\right|_{\rm CE,\,aperture}} = {\left. {R_{\rm aperture}\frac{{dN}}{{d{E'}}}} \right|_{\rm CE}},
\nonumber
\end{equation}
where $R_{\rm aperture}$ is the {\it reduction factor}
\begin{multline}
R_{\rm aperture}=\frac{1}{2\pi {{\sigma }_{{{f,x}}}}{{\sigma }_{{{f,y}}}}}\\
\iint\limits_{\rm aperture}{\exp \left( -\frac{x_a^2}{2\sigma _{{{f,x}}}^{2}} \right)}\exp \left( -\frac{y_a^2}{2\sigma _{{{f,y}}}^{2}} \right)d{x_a}d{y_a},\label{eq:reductionfactor}
\end{multline}
which unambiguously determines the peak height of the spectrum in terms of the total number density of the Compton-edge (CE) photons
\begin{equation}
{\left. \frac{{dN}}{{d{E'}}} \right|_{\rm CE}}=\frac{3N_{\rm tot}}{2 E_{\gamma,{\rm max}}}
\nonumber
\end{equation}
integrated over the whole aperture plane \cite{Sun2011}. The quantity $N_{\rm tot}$ gives the total number of scattered photons in the aperture plane as defined in Section \ref{sec:x-ray}.
When all the scattering angles for the aperture are much smaller than the spreads, the exponentials are both 1 throughout the integration region and one obtains the correct
\begin{equation}
R_{\rm aperture}=\frac{{{A}_{\rm aperture}}}{2\pi {{\sigma }_{{{f,x}}}}{{\sigma }_{{{f,y}}}}}.
\label{eq:R_small_aperture}
\end{equation}
In the other limit $R_{\rm apertrure}=1$, as all the electrons contribute to the edge upon integrating over the full aperture plane.
When ${{\sigma }_{{{f,x}}}}={{\sigma }_{{{f,y}}}}=\sigma_f$ and the aperture is circular with a radius $r_a$, the integral evaluates to
\begin{equation} \label{eq_reduction}
R_{\rm aperture}=1-\exp \left( -\frac{r_{a}^{2}}{2\sigma _{f }^{2}} \right).    
\end{equation}

\begin{figure}[h!]
\centering\includegraphics[width=\linewidth]{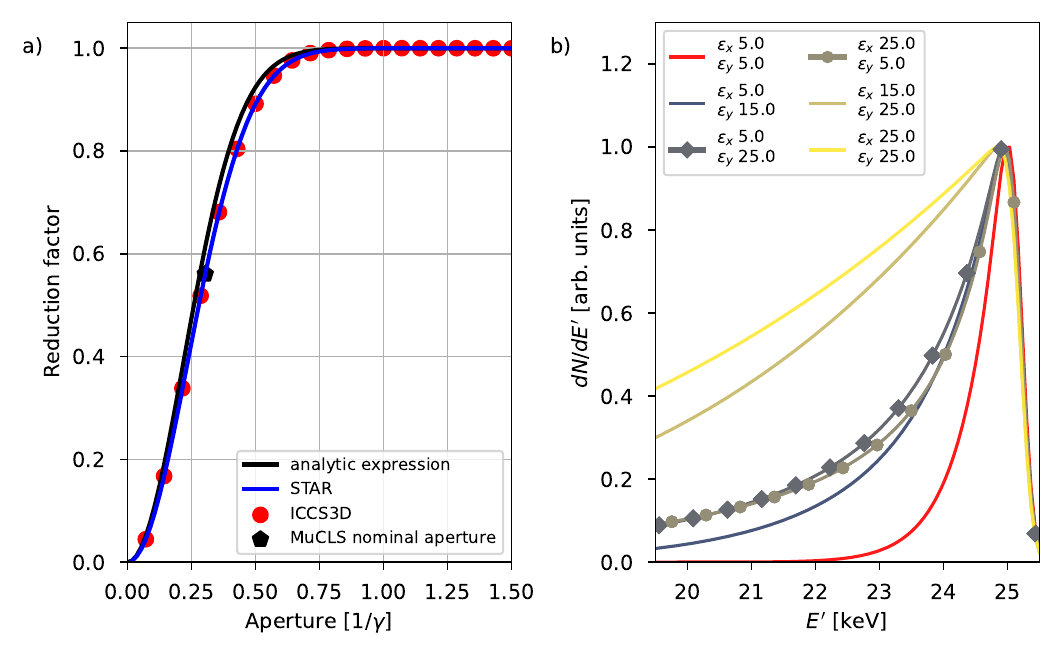}
\caption{
a) demonstrates the agreement of the presented theory with the STARS-code based on this theory and ICCS3D, a state-of-the-art code for Compton simulation.
Typical values for the MuCLS, the ICS employed in this study, were used for this comparison.
Except for the normalized emittances $\varepsilon_{x}=9.3\ \si{\mm \milli\radian}$ and $\varepsilon_{y}=11.1\ \si{\mm \milli\radian}$, all parameters are displayed in Table \ref{tab:pulses}.
The analytic prediction of Eq.~(\ref{eq_reduction}) is a good approximation even in the case of slightly asymmetric beam diameters and emittances.
In b), the effect of the electron beam emittances on the X-ray spectra is visualized for the laser and electron beam parameters of the MuCLS.
These graphs represent the the X-ray spectrum as produced at the interaction point.
The small emittance dominates close to the peak, while the low-energy fall-off is described by the larger one.}
\label{fig:reduction_final}
\end{figure}

Figure \ref{fig:reduction_final} a) shows the reduction factor for the spectrum's peak returned by the analytical formula and full simulations of the radiation passing through a circular aperture for the nominal MuCLS parameters given in Table \ref{tab:pulses} and varying aperture.

\subsection{The X-ray distribution fall-off length}

Next, the distribution of scattering angles for photons passing through a small aperture in the forward direction at $x_a=0$ and $y_a=0$ is computed and the spectrum is analysed following the statistical argument in \cite{ranjan18}.
The propagation equations going back to the source location are now
\[\theta_x=-\theta''_x-x_s/L\]
and
\[\theta_y=-\theta''_y-y_s/L,\]
where $\theta''_x=\theta'_x-\theta_x$ and $\theta''_y=\theta'_y-\theta_y$ are the transverse photon scattering angles relative to the electron angles.
They vanish for Compton-edge photons.

The resulting unit normalized distribution is
\begin{multline} \label{eq:mult_falloff_length}
f(\theta''_x,\theta''_y;x_a=0,y_a=0)=\\
\int \int \frac{\exp(-x_s^2/2\sigma^2_{s,x})\exp(-y_s^2/2\sigma^2_{s,y})}{2\pi\sigma_{s,x}\sigma_{s,y}}\\
	\frac{\exp(-(\theta''_x+x_s/L)^2/2\sigma_{\theta_x}^2)
	\exp(-(\theta''_y+y_s/L)^2/2\sigma_{\theta_y}^2)}{2\pi
	\sigma_{\theta_x}\sigma_{\theta_y}}dx_sdy_s\\
	= \frac{L^2\exp(-{\theta''_x}^2L^2/2\sigma^2_{f,x})\exp(-{\theta''_y}^2L^2/2\sigma^2_{f,y})}{2\pi\sigma_{f,x}\sigma_{f,y}}.
\end{multline}

As a function of the photon scattering angle, the energy of the scattered radiation is given by the Thomson formula as, e.g.\ \cite{Sun2011,krafft16,Arutyunian1963},
\begin{equation}
{E}'={{E}_{p}}\frac{1+\beta }{1-\beta \cos {\theta }''}\approx {{E}_{p}}\frac{{{\left( 1+\beta  \right)}^{2}}{{\gamma }^{2}}}{1+{{\gamma }^{2}}{{{{\theta }''}}^{2}}}
\label{eq:CompEdge}
\end{equation}
with  ${\theta }''=\sqrt{{\theta }_{x}^{\prime\prime 2}+{\theta }_{y}^{\prime\prime 2}}$.
When the source distribution and the distribution in electron angles are such that ${{\sigma }_{{_{f,x}}}}={{\sigma }_{{{f,y}}}}={{\sigma }_{f }}$, the distribution in scattering angles may be converted into a distribution in scattered energy by performing the change of variables $\theta_x'', \theta_y'' \to \theta'',\phi$, with an integration over $\phi$ and the second transformation ${{{\theta }''}^{2}}\to {E}'$: 
	\begin{align*}
  {{{{\theta }''}}^{2}} & =-\frac{{E}'-{{E}_{\gamma ,\max }}}{{{\gamma }^{2}}{{E}'}}, \\ 
 \frac{df}{d{{{{\theta }''}}^{2}}} & =-\frac{L^2}{2\sigma _{f }^{2}}\exp \left( -\frac{{\theta''^2}}{2(\sigma _{f }^{2}/L^2)} \right), \\ 
 \frac{df}{d{E}'} & =\frac{df}{d{{{{\theta }''}}^{2}}}\frac{d{{{{\theta }''}}^{2}}}{d{E}'} \\ & =\frac{L^2{E}_{\gamma ,\max }}{2{{\gamma }^{2}}\sigma _{f }^{2}{{E}'^2
 }}\exp \left( \frac{{E}'-{{E}_{\gamma ,\max }
 }}{2{{\gamma }^{2}}(\sigma _{f }^{2}/L^2){{E}'}} \right)\Theta \left( {{E}_{\gamma ,\max }}-{E}' \right),
\end{align*}
where $\Theta \left(x \right)$ is the step function, $\Theta \left(x \right)=1$ for $x\geq0$ and $\Theta \left(x \right)=0$ for $x<0$.
Both, the ${{{\theta }''}^{2}}$ distribution and the ${E}'$ distribution are unit normalised.\\
This distribution in the small aperture limit captures the main qualitative and quantitative features for larger apertures.
First, no photons are emitted beyond the Compton-edge energy of $E_{\gamma,{\rm max}}$.
Second, the peak value of the distribution occurs when $E'=E_{\gamma,{\rm max}}$, that is, at the Compton-edge energy.
Third, an exponential falloff on the low-energy side of the distribution is present.
Clearly, the energy falloff length for equal effective source sizes  is
\[l_{\rm falloff}=2\gamma^2(\sigma_f^2/L^2)E_{\gamma,\max}.\]
Such an exponential falloff length, whose value increases with the aperture area, is present for large apertures \cite{Krafft2026}.
In the general case of unequal effective source sizes, two falloff lengths are present, cf.\ Eq.~(\ref{eq:mult_falloff_length}).
Near the Compton edge, the falloff length is close to the one of the dimension with the smaller spread.
In the tail of the distribution, the falloff length of the dimension with the larger spread dominates.
Figure \ref{fig:reduction_final} b) depicts this effect.
The symmetry of the $\varepsilon_{x}=5\ \si{\mm \milli\radian}$ \& $\varepsilon_{y}=25\ \si{\mm \milli\radian}$ curve and the $\varepsilon_{x}=25\ \si{\mm \milli\radian}$ \&  $\varepsilon_{y}=5\ \si{\mm \milli\radian}$ one is lifted due to the asymmetric electron beam focus size.

\subsection{The X-ray spectrum's normalisation}
\label{sec:x-ray}

The overall normalisation of the scattered photon spectral density curve may be found by considering a special case.
For zero emittance, all electrons are aligned with the $z$-axis and they radiate energy at the Compton edge in the forward direction.
All the scattered radiation at the edge energy is found by integrating over the total aperture plane leading to $R=1$.
The total $dN/d{E}'$  in the forward direction is the single electron contribution multiplied by the number of electrons.
For aligned electrons, the radiation from a single electron averaged over the transverse distribution of the photon pulse is \cite{Sun2011}
	\[{{\left. \frac{dN}{d{E}'} \right|}_{1\text{ } \rm electron, CE }}=\frac{3}{2}\frac{{{N}_{\rm tot}}}{{{N}_{e}}}\frac{1}{{{E}_{\gamma ,\max }}},\]
with

\begin{align*}
{{N}_{\rm tot}} & =\frac{{{\sigma }_{T}}}{2\pi \sqrt{\sigma _{e,x}^{2}+\sigma _{p,x}^{2} }\sqrt{\sigma _{e,y}^{2}+\sigma _{p,y}^{2} }}{{N}_{e}}{{N}_{p}} \\
& =\frac{8\pi r_{e}^{2}}{3}\frac{{{N}_{e}}{{N}_{p}}}{2\pi \sqrt{ \sigma _{e,x}^{2}+\sigma _{p,x}^{2} }\sqrt{ \sigma _{e,y}^{2}+\sigma _{p,y}^{2} }},
\end{align*}
where $N_e$ and $N_p$ are the number of electrons and photons, respectively;
$\sigma_T$ is the Thomson cross-section, $r_e$ is the classical electron radius, and $\sigma_e$ and $\sigma_p$ are the transverse rms sizes of the electron beam and laser pulse intensity, respectively.
To get agreement with the Compton-edge value for the spectral density in this specific case the normalized distribution must be multiplied by the proper constant factor.

To derive the constant, consider the case of a small circular aperture in the forward direction to the electron beam.
For small forward scattering angles, the differential scattering cross section integrated over scattering angle $\phi$ has the value \cite{Tomassini2005}
\begin{align*}
\frac{d\sigma}{d\Omega} & =\frac{r_e^2}{\gamma^2(1-\beta\cos\theta'')^2}\left[\frac{1}{2}+\frac{(\beta-\cos\theta'')^2}{2(1-\beta\cos\theta'')^2}\right] \\
& =\frac{E'^2}{\gamma^2
E_p^2(1+\beta)^2}\frac{r_e^2}{{2{\beta ^2}}}\left[ {{\beta ^2} + {{\left( {\frac{{\left( {1 + \beta } \right){E'}}}{{{E_{\gamma ,\max }}}} - 1} \right)}^2}} \right].
\end{align*}
Within a small aperture ($r_a \ll \sigma_f$) subtending a solid angle $d\Omega$, the emission is close to the Compton-edge energy and the total number of photons collected in the aperture is
\begin{align*}
N_{\rm aperture}&=\frac{1}{\sigma_T}\frac{d\sigma}{d\Omega}d\Omega N_{\rm tot} \\
&=\frac{(1+\beta)^2\gamma^2r_e^2}{8\pi r_e^2/3}\frac{\pi r_a^2}{L^2} N_{\rm tot} =\frac{3}{2}\gamma^2\frac{r_a^2}{L^2}N_{\rm tot}.
\end{align*}
The normalization result is proportional to the aperture area as it should be.
In this same limit the overall spectral density must be
\begin{multline}
{{\left. \frac{dN}{d{E}'} \right|}_{\rm aperture}}=N_{\rm aperture}\frac{df}{dE'} \\
=\frac{3}{2}\frac{{N}_{\rm tot}{E_{\gamma,\max}}}{E'^2}
 R_{\rm aperture}\\
 \exp \left( \frac{{E}'-{{E}_{\gamma ,\max }
 }}{2{{\gamma }^{2}}(\sigma _{f }^{2}/L^2){{E}'}} \right)\Theta \left( {{E}_{\gamma ,\max }}-{E}' \right).
 \label{eq:fund_sol}
\end{multline}
In this case, $R_\textrm{aperture}= \pi r^2_a / (2\pi \sigma^2_f)$, which is exactly the small aperture limit, Eq.~(\ref{eq:R_small_aperture}).
It is obvious that this fundamental solution
integrated in $E'$ yields $N_{\rm aperture}$.
This formula applies very well in more general situations in which the emittances or the source sizes are not small, but does not cover the general case $\sigma_{f,x}\ne\sigma_{f,y}$.

For the general case, and at a specific value of $E_p$ and $\gamma$, the reduction factor is
\begin{align} \label{eq_4fromotherpaper}
& R(E')  = \frac{1}{{2{\beta ^2}}}\left[ {{\beta ^2} + {{\left( {\frac{{\left( {1 + \beta } \right){E'}}}{{{E_{\gamma ,\max }}}} - 1} \right)}^2}} \right]\Theta \left( {{E}_{\gamma ,\max }}-{E}' \right)
\nonumber \\
& \times \frac{1}{4\pi^2\sigma_{f,x}\sigma_{f,y}}\iint\limits_{\mathrm{aperture}}
\int\exp \left( { - \frac{{{{\left( {{x_a} - \sin \theta ''\cos \phi' L} \right)}^2}}}{{2\sigma _{{f,x}}^2}}} \right)
\nonumber \\
& \times \exp \left( { - \frac{{{{\left( {{y_a} - \sin \theta ''\sin \phi' L} \right)}^2}}}{{2\sigma _{{f,y}}^2}}} \right) d{x_a}d{y_a}d\phi',
\end{align} 
where the scattered angle is determined by solving the scattered energy relation Eq.~(\ref{eq:CompEdge}) for $\sin \theta ''$.
The first two factors in Eq.~(\ref{eq_4fromotherpaper}) capture the dependence of the differential scattering cross section on the scattering angle.
The third factor, as above, ensures there is no emission beyond the Compton edge.
The factors within the integral provide the correct generalization of Eq.~(\ref{eq:reductionfactor}), including the scattered energy dependence of the spectrum implicit in Eq.~(\ref{eq:CompEdge}). Indeed, the expression is identical to Eq.~(\ref{eq:reductionfactor}) for photons scattered with the Compton-edge energy when $\sin \theta ''=0$.

The high-energy edge of the spectrum is much sharper than the one on the low-energy side.
Its width is determined by a combination of the relative energy spreads of the electron beam and the incident laser beam.
It can be calculated by simply integrating the fundamental solution over the two normalized energy distributions

\begin{equation} 
{{\left. \frac{dN}{d{E}'} \right|}_{\rm aperture}}=\frac{3}{2}{N}_{\rm tot}
 \times\int\int \frac{n_\gamma(\gamma)n_p(E_p)}{E_{\gamma,\max} }R(E')d\gamma dE_p,
 \label{eq:spectrum_shape}
 \end{equation}
where now ${E}_{\gamma, \max}$ is defined for each value of the integration variable for the photon energy distribution $E_p$ and electron energy distribution $\gamma$.

\subsection{Implementation of the quantitative model into the fast {\tt STARS} code for Compton scattering simulations}\label{subsec:star_algrithm}

Equation (\ref{eq:spectrum_shape}) forms the basis of the quantitative model that was implemented into the computer code {\tt STARS}: {\tt S}tatistical {\tt T}reatment of {\tt A}dvanced {\tt R}adiation {\tt S}pectra.
The code is written in Python.
The only aspect of a realistic simulation of (inverse) Compton scattering not captured by the quantitative model of the previous section is electron energy spread because the analysis is carried out for a single energy. 
{\tt STARS} incorporates the electron energy spread by convolving spectra due to electron energies drawn from an arbitrary distribution.
However, this does not incur significant additional computational cost because spectra emitted by electrons with nearby energies are only shifted versions of each other (see Section \ref{sec_shape_spectra}). 
Due to the relative simplicity of the integration in the physical model and this convenient scaling property, {\tt STARS} is orders of magnitude faster than our earlier code {\tt ICCS3D} \cite{krafft16,ranjan18}.
Appendix \ref{app:comp_STARS} validates that STARS yields the same results as {\tt ICCS3D} and CAIN \cite{CAIN}.

\section{Strategy for the complete characterisation of an inverse Compton X-ray source}\label{sec:characterisation_strategy}

Since diagnostic capabilities are usually very limited at compact ICSs, especially inside small storage rings, the strategy proposed in the following retrieves the electron beam parameters from information encoded in the X-ray beam parameters when those of the laser beam are known.
This is possible because most laser and X-ray beam parameters can be determined directly.
The {\tt STARS} simulation code is employed to fit the electron beam emittance such that the computed X-ray spectrum matches the measured one best.\\
In order to do so, the following characterization measurements are performed to determine the required parameters:
\begin{enumerate}[nosep]
    \item Characterisation of the interaction laser
        \begin{enumerate}[nosep]
            \item Determination of the interaction angle by measuring the laser beam orbit tilt with respect to the electron beam orbit.
            \item  Determination of the interaction laser's pulse length. In our case, an autocorrelator is employed for this purpose, cf.\ \cite[Chapter 7.2.2]{Gunther2023}.
            \item Determination of the laser focal spot size from a measurement of the transverse mode range, cf.\ \cite[Chapter 7.3.2]{Gunther2023}.
            \item Ringdown measurement to retrieve the laser power, 
             pulse energy and number of photons \cite{Rempe1992,An1995,Virgo2007}.
        \end{enumerate}
    \item Characterisation of the X-ray beam
        \begin{enumerate}[nosep]
            \item Measurement of the X-ray spectrum with a silicon drift detector.
            \item Determination of the transverse X-ray sources size from a knife-edge measurement \cite{Gunther2019}.
            \item Retrieval of the total X-ray photon flux with a calibrated CCD camera \cite{Gunther2019}.
        \end{enumerate}
\end{enumerate}
The cited literature provides more details on the mentioned procedures.
Once these parameters are known, the missing electron beam parameters are retrieved in the following manner:
\begin{enumerate}[nosep]
    \setcounter{enumi}{2}
    \item Characterisation of the electron beam\newline
        Inferring the electron beam parameters requires an iterative procedure. First, the start parameters are estimated.
        \begin{enumerate}[nosep]
            \item Determine the electron beam pulse length. In our case, this was done by the manufacturer of our ICS, cf.\ \cite{Schleede2013a}.
            \item Calculate an initial guess for the electron beam divergence based on the nominal symmetric design parameters for the emittance of \qty{12}{\mm \milli\radian} and the electron focus size of \qty{50}{\um}.
            \item Estimate the electron beam focal spot size by matching the projected X-ray source size calculated from the spatial overlap of the laser and electron beam to the measured X-ray spot size.
            The electron beam size is varied while keeping the divergence constant.
            A Gaussian profile is assumed for both beams.
            \label{item:ebeam_focus}
        \end{enumerate}
    After the start values for these electron beam parameters are calculated, the electron beam size and emittance as well as its mean energy and energy spread are retrieved by iterating between a genetic algorithm updating the electron beam emittance, i.e.\ its divergence, and the electron beam size determination based on the projected X-ray source size, cf.\ step \ref{item:ebeam_focus}.
    This is necessary because both electron beam parameters affect the shape of the X-ray spectrum, cf.\ Figures  \ref{fig:reduction_final} b) \& Appendix \ref{app:influence_emittance_focus}.
        \begin{enumerate}[nosep]
            \setcounter{enumii}{3}
            \item Retrieve the electron beam emittance which gives the best fit between simulated and measured spectrum by using a genetic algorithm running {\tt STARS} with different emittance values.
            In addition, the electron beam's mean energy and its energy spread are allowed to vary.
            \label{item:emittance_det}
            \item Calculate the new electron beam focus based on the updated emittance from step \ref{item:emittance_det} like in step \ref{item:ebeam_focus}.\label{item:ebem_focus_update}
            This is necessary because a change of the emittance affects the beam divergence and hence the projected X-ray source size, cf.\ Appendix \ref{app:influence_emittance_focus}.
            \item Repeat steps \ref{item:emittance_det} and \ref{item:ebem_focus_update} until convergence is reached. \label{item:iteration_end} 
            \item Calculate the X-ray flux with {\tt STARS} using all the parameters determined in the previous steps.
            \label{item:flux}
        \end{enumerate}
\end{enumerate}

\section{The Munich Compact Light Source (MuCLS)}
The strategy presented in the preceding chapter was employed at the Munich Compact Light Source (MuCLS) to retrieve its electron beam properties.
The MuCLS is a compact synchrotron hard X-ray facility based on a storage ring-based inverse Compton X-ray source \cite{Gunther2023}.
In the following, the main components of the MuCLS are described, especially those relevant for this study.
The MuCLS facility consists of three subsystems, namely the electron beam system, the interaction laser system and the X-ray beamline.
All components of the inverse Compton X-ray source are synchronised to a master frequency operating at \SI{2856}{\mega\hertz}.

\subsection{The electron beam system}
The electron beam of \qty{\sim 250}{\pico \coulomb} charge is generated from a Copper-photocathode illuminated with a \qty{\sim 30}{\pico \second} (full-width at half maximum) laser beam.
Afterwards, the electrons are accelerated and injected into a compact storage ring of \qty{4.6}{\m} circumference, corresponding to a repetition rate of \qty{64.91}{\MHz}.
The energy of the stored electron beam can be tuned between \qtyrange[range-phrase= { and }]{29}{45}{\MeV} via the linear accelerator.
For this study, the electron energy was set to \qty{37.6}{\MeV} to produce X-rays with a peak energy of \qty{\approx25}{\keV}.
The electron bunch is focused at the collision point with the laser beam.
The electron beam's pulse length is \qty{50}{\pico \second} \cite{Schleede2013a}.
One bunch is circulating at a time inside the storage ring in a non-equilibrium condition to preserve a round transverse beam profile.
After \qty{40}{\milli \second}, i.e.\ at a rate of \qty{25}{\Hz}, the circulating bunch is replaced with a new one.

\subsection{The interaction laser system}\label{subsec:int_las_syst}
The interaction laser system providing the infrared (IR) photons for the ICS process consists of two main devices.
An all solid-state pico-second Nd:YAG laser system ($\lambda= 1064~\mathrm{nm}$, repetition rate \qty{64.91}{\mega\hertz}) delivers an average optical power of about \qty{30}{\watt}.
The M\textsuperscript{2}-factor of its laser pulses is \textrm{1.3}.
These are injected into a high-finesse laser resonator, also called enhancement cavity.
It amplifies the laser pulse energy by four orders of magnitude and recirculates the laser pulse similarly to a storage ring.
The laser's propagation direction inside the enhancement cavity is designed such that the laser pulse counter-propagates the electron beam in their common straight section and incidents onto the latter at a small angle of about \qty{6}{\milli \radian}, cf.\ Appendix \ref{app:interaction_angle}.
This justifies assumption \ref{item:head_on}.
The systematic downshift in X-ray energy by such an interaction angle is negligible.
Since the resonator length is \qty{9.2}{\m}, its repetition rate is \qty{32.46}{\MHz}, i.e. the \textrm{88}\textsuperscript{th} sub-harmonic of the master frequency.
Accordingly, two laser pulses are stored inside the resonator at the same time.
The solid-state laser is locked to the enhancement cavity via the Pound-Drever-Hall locking mechanism \cite{Pound1946,Drever1983,Black2001}.
The cavity is locked to the master frequency via a second feedback, thereby ensuring synchronisation to the electron beam.
The resonator design is an all-curved mirror bow-tie cavity containing three tight foci.
The resonator geometry is displayed in Figure \ref{fig:TMR_scan} a).
Off-axis incidence results in astigmatism.
Since a single mirror compensates the round-trip effect, the laser beam becomes slightly elliptic, cf.\ Figure \ref{fig:TMR_scan} b).
A detailed description of the interaction laser system and the enhancement cavity can be found in \cite[Chapter 7]{Gunther2023}.

\subsection{X-ray beamline}
Compton scattering of the laser photons at the electrons produces the X-ray radiation.
Interaction of both beams at their common waist is ensured by tuning the X-ray beam for maximum flux at minimum X-ray source size.
Therefore, assumption \ref{item:alpha_eq_zero} holds, which is supported in Appendix \ref{app:alignment_spectrometer}.
At \qty{350}{\kilo\watt} of stored laser power and \qty{250}{\pico\coulomb} charge, a brilliance of \qty{1.2e10}{\photon\per \second \per \milli\metre\squared \per \bw} is reached.
X-ray photons emerge from the Compact Light Source (CLS) through a fixed-diameter aperture which confines the X-ray beam to a divergence of approximately \qty{4}{\milli\radian} \texttimes\, \qty{4.5}{\milli\radian} (horizontal \texttimes\, vertical).
The X-rays exit the CLS through a thin Beryllium vacuum window and propagate towards the experimental setups.
The first setup is located directly downstream of the radiation shielding enclosure of the CLS.
It contains a very flexible sample and detector mounting system that allows for a minimum X-ray source-to-sample distance of \qty{3.5}{\meter} and a maximum one of \qty{5.2}{\meter}.
Hence, this setup was employed for the X-ray spectrum measurements.
More details on the X-ray beamline can be found in \cite{Gunther2020}.

\section{Characterisation of the laser enhancement cavity}\label{sec:ec_characterisation}

\subsection{Guoy-phase measurement for laser focus evaluation}\label{subsec:las_foc_det}
\begin{figure}
    \centering
    \includegraphics[width = \linewidth]{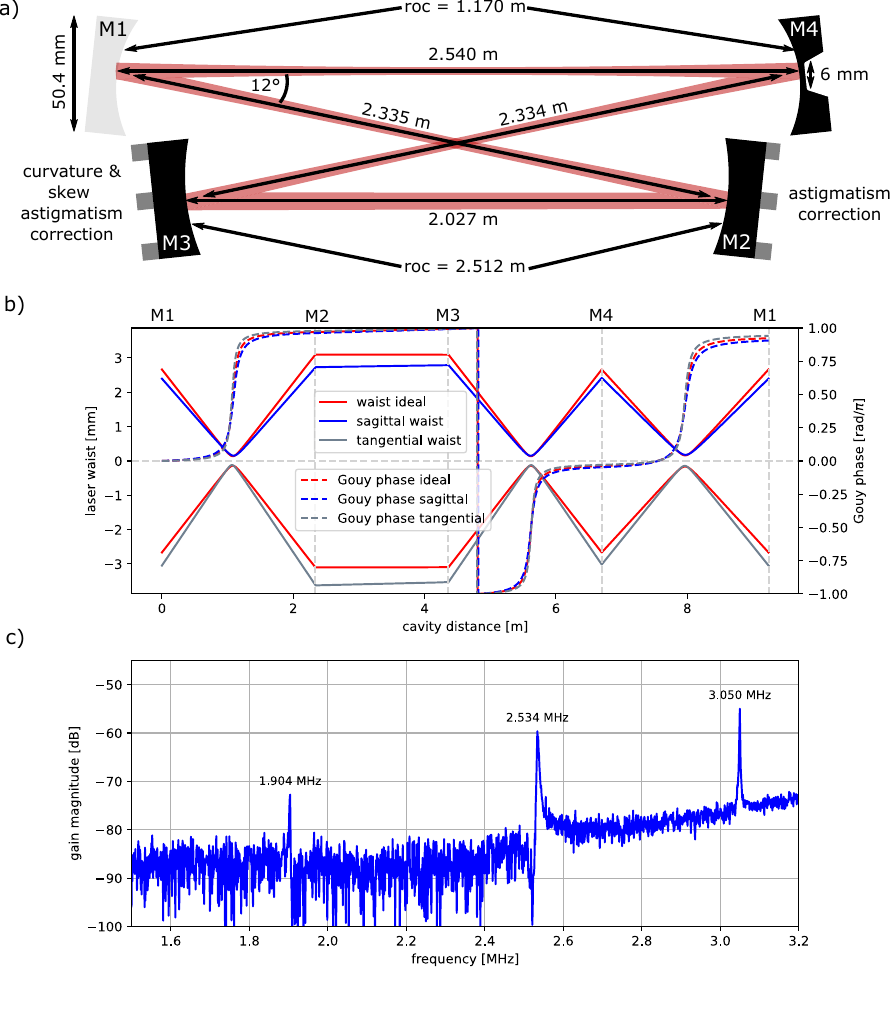}
    \caption{Key features of the MuCLS enhancement cavity.
    In a), the geometry of the enhancement cavity is drawn, see also \cite[Chapter 7]{Gunther2023} for more details.
    The numbering of the mirrors M1 to M4 indicates the propagation direction.
    The abbreviation roc denotes the mirror radius of curvature.
    The laser mode waist and the Gouy phase for the cavity depicted in a) is displayed in b).
    In the ideal case, no angle of incidence is considered, while the tangential and sagittal cases correspond to the cavity's thermal state for which the transverse mode range shown in c) was recorded.
    From this measurement, the resonator's tangential waist of \qty{143.2}{\um} with a corresponding Rayleigh length of \qty{6.05}{\cm} is retrieved.
    The corresponding sagittal ones are \qty{179.8}{\um} and $z_{R}=$  \qty{9.54}{\cm}.
    Panel a) is adapted from Figure 7.8 a) in \cite{Gunther2023}.
    }
    \label{fig:TMR_scan}
\end{figure}

The frequency spacing of higher-order transverse modes to the fundamental one is correlated to the round-trip Gouy phase in the enhancement cavity.
Measuring a cavities transverse mode spectrum, the Gouy-phase can be inferred, e.g., demonstrated by \cite{Stochino2012,Djevahirdjian2020}. 
Accordingly, this diagnostic can be applied to determine a mirror's radius of curvature \cite{Uehara1995,Uehara1995a} or retrieve a cavity's thermal state \cite{Mueller2015}.
The diagnostic implemented at the MuCLS works very similar to those mentioned above and is described in \cite[Chapter 7.3.2; pp.\ 164-166]{Gunther2023}.\\
The higher-order mode with a mode sum of $2$, located about \qty{3}{\mega\hertz} below the fundamental one, is used for determining the Rayleigh length of the enhancement cavity.
This is achieved by sweeping the sidebands between \qtyrange[range-phrase={ and }]{1.5}{3.5}{\mega\hertz} because the resonance of this TEM is expected in this range.
The network analyser swept the frequency across the \qty{2}{\mega\hertz}-span in $2048$ steps.
Its receiver bandwidth was set to \qty{1}{\kHz}.
Three distinct resonances appear in the spectrum of the transverse-mode range that correspond to the three modes TEM\textsubscript{20}, TEM\textsubscript{11}, and TEM\textsubscript{02}.
The process for the retrieval of the tangential and sagittal Rayleigh lengths and focal waists from the measured frequency spacing is described in \cite[Chapter 7.3.2; pp.\ 166-167]{Gunther2023} and Appendix \ref{app:thermal_effects}.
The retrieved tangential waist is \qty{143.2}{\um} and the sagital one is \qty{179.8}{\um} corresponding to Rayleigh length of \qty{6.05}{\cm} and \qty{9.54}{\cm} for the spectrum depicted in Figure \ref{fig:TMR_scan} c).
Figure \ref{fig:TMR_scan} b) depicts the resulting laser beam waist for one round-trip inside the enhancement cavity.

\subsection{Laser pulse length determination}\label{subsec:las_puls_len}
The temporal duration of the laser pulse determines the length of the ``laser undulator'' and thus influences both the X-ray flux and spectrum.
However, measuring the absolute pulse length inside the enhancement cavity is impossible at the MuCLS.
Nevertheless, a very good estimate can be obtained by measuring the laser pulse duration right in front of the enhancement cavity since the cavity itself is in ultra-high vacuum.
Accordingly, the only dispersive element is the thin input coupling mirror which the laser has to traverse only once.
Since an Nd:YAG laser is used at the MuCLS, its natural pulse duration is typically limited to tens of picoseconds which allows to neglect the input coupling mirror's dispersion.
Further, an autocorrelation measurement is well suited to determine laser pulse durations in the picosecond range.
An FR-103XL (Femtocrome Research Inc., Berkeley, USA) autocorrelator was employed for this task.
The retrieved pulse duration was \qty{11.2}{\pico\second} (rms).
More details can be found in \cite[section 7.2.2]{Gunther2023}.

\subsection{Stored optical power}
The optical power stored inside the enhancement cavity is retrieved from a ringdown measurement \cite{Rempe1992,An1995,Virgo2007}.
The input coupler's transmission has to be known for this technique.
It was determined to be \qty{124}{ppm} by the manufacturer of the CLS, Lyncean Technologies Inc.; measurement uncertainties were not provided.
Furthermore, the free-spectral range must be known.
Since the enhancement cavity is locked to an external reference, a sub-harmonic of the \qty{2856}{\MHz} master clock, the free-spectral range is known very precisely.
Hence, this parameter does not introduce inaccuracies to the measurement.
Details on the measurement can be found in \cite{Loewen2003} and \cite[Chapter 4.2.5 \& 7.3.3]{Gunther2023}.
The stored power during this experimental campaign is about \qty{105}{\kW}.
The laser pulse's energy can be calculated by dividing the stored power by the number of interactions, i.e., the product of the cavity's free spectral range and the number of pulses stored therein (two pulses in our case).
Once the pulse energy is known, the number of photons $N_{\rm tot}$ is retrieved dividing the pulse energy by the photon energy of a $\lambda=1064~\si{\nm}$ Nd:YAG laser photon.\\

\section{X-ray parameter determination}
\subsection{X-ray source parameter determination}
X-ray beam parameter determination was performed with the X-ray beam monitoring system described in detail in \cite{Gunther2019}.
The image of a knife-edge intercepting a small part at the very bottom of the X-ray beam is recorded with an X-ray detector based on a fibre-coupled scintillator and CCD technology.
The point-spread function of the camera was determined and is deconvolved from the image.
As the blur of the knife-edge in the image is caused by the extended source size, the horizontal and vertical source parameters (size and position) can be extracted from an error function fitted to the respective edges if the geometry of the setup is known.
The horizontal and vertical X-ray source size were determined to be \qty{42.6}{\um} and \qty{46.3}{\um}, respectively, during this measurement campaign.
The X-ray flux is determined by averaging the number of counts in a small area and extrapolating it to the full size of the X-ray beam.
Translation of the camera counts to quantitative X-ray photon numbers is realised by cross-calibrating the camera to a photon-counting detector for each X-ray energy setting of the MuCLS.
At \SI{25}{\keV} X-ray energy, \SI{0.73e10}{ph/s} exited the CLS through the Beryllium vacuum window during the experiment presented here.
Consequently, the projected X-ray source size, position, and the X-ray flux can be acquired in parallel to experiments using this X-ray beam monitoring system.

\subsection{X-ray spectrum measurement}
In addition to the X-ray source size, the emitted X-ray spectrum must be known for our approach.
This subsection presents the acquisition of the X-ray spectra, explains the contributions modifying the X-ray spectrum during beam propagation from the X-ray source to the X-ray spectrometer, and discusses how these modulations are accounted for in the analysis and electron beam parameter retrieval.
The energy-dependent absorption of any material between the X-ray source point and the X-ray spectrometer, in particular the spectrometer detector material itself, alters the measured X-ray spectrum.
Consequently, the measured spectrum is modified by the following effects compared to the spectrum actually generated at the interaction point:
\begin{itemize}
    \item energy-dependent absorption of the X-rays due to propagation in air and other materials on their path to the spectral detector (beam hardening);
    \item the energy-dependent absorption of the X-ray photons in the X-ray spectral detector (opposite effect of beam hardening).
\end{itemize}
Aside from air, the other objects inside the X-ray beam path are the back-thinned laser mirror, the beryllium vacuum window of the laser resonator, mylar foils at each end of the vacuum beampipe, and a \qty{13.95}{\mm} thick aluminium block which intentionally reduces the X-ray flux to a count rate acceptable by the silicon-drift-detector (SDD) of the spectrometer.
Absorption by the thin beryllium window protecting the SDD of the spectrometer from visible light has to be included in the calculation of beam hardening as well.
The energy-dependent transmission of these materials is calculated and combined with the absorption of the silicon-drift-detector into a single ``energy-dependent detector absorption function''.
Since {\tt STARS} calculates the spectra at the interaction point, its spectrum must be multiplied with this ``energy-dependent detector absorption function''.
The resulting corrected curves are compared to the measured spectra in the analysis of the electron beam parameters.\\
The spectra were recorded with an AXAS-D SDD-spectrometer with a VITUS H20 (KETEK GmbH, Germany).
The detector was aligned at the centre of the radiation cone at a distance of \qty{5}{\m} downstream the X-ray source point, justifying assumption \ref{item:aperture_on_axis}.
The alignment is described in Appendix \ref{app:alignment_spectrometer}.
The acquisition time for one spectrum was set to \SI{120}{\s} to accumulate sufficient statistics even though the aluminium block was placed into the X-ray path.
Since the recorded spectrum is subject to Poisson statistics, it was smoothed with a boxcar moving average filter, taking into account $21$ neighbouring points ($10$ points before and $10$ points after the one in question).
The simulation is fitted to these smoothed measured spectra.

\section{Determination of the electron beam parameters} \label{Sec_BeamParams}
\subsection{Matching simulated and measured spectra: an optimisation problem}
\label{sec:GA}
The simulation code {\tt STARS} introduced in subsection \ref{subsec:star_algrithm} was employed for the spectrum calculation.
It is employed in steps \ref{item:emittance_det} and \ref{item:flux} in section \ref{sec:characterisation_strategy}.
In order to use {\tt STARS} as a diagnostic tool, a multi-dimensional, non-linear optimisation problem is formulated in which the least square difference between the experimental data, i.e., the spectral density $dN/d{E}'|_e$, and the corresponding numerical values, $dN/d{E}'|_n$, 
computed with {\tt STARS} is minimised:
\begin{equation}
    \chi^2 = \sqrt{\frac{1}{M} \sum_{i=1}^{M} \left(\left.\frac{dN}{d{E}'_i}\right|_e-\left.\frac{dN}{d{E}'_i}\right|_n\right)^2},
\end{equation}
at some set of $M$ scattering energies $E_i'$.
A genetic algorithm optimises the $\chi^2$ fit between experimental data and simulations over the four electron beam parameters: mean energy, energy spread, horizontal and vertical emittances. 
We use automation systems that build on the Platform and Programming Language Interface for Search Algorithms ({\tt PISA}) developed at ETH Z\"urich \cite{Bleuler2002a} and Alternate PISA ({\tt APISA}) from Cornell University \cite{bazarov2005}.
The codes are changed to enable efficient job execution and management on various (parallel) computational platforms, as well as to easily plug in different ``function evaluators''---simulation software programs which carry out numerical simulations. 
In this study, that is {\tt STARS}.
For a detailed and pedagogical overview of the genetic algorithm used, see \cite{hofler2013}. 
Genetic algorithm optimisation dispatches a single ``generation'' of simulations at a time, each with different values for the parameters to be optimised (sampling predefined ranges). 
When they are all completed, the $\chi^2$ values of each of these ``individuals'' is used to rank them and use them to compute the parameters' values for the next generation.
By nature, these optimisations are ideal for parallel platforms, such as computer clusters.
However, they can also run on a powerful, stand-alone workstation.
The optimisation is performed in two consecutive steps which are described in the following two subsections.

\begin{figure}[h!]
\centering\includegraphics[width=\linewidth]{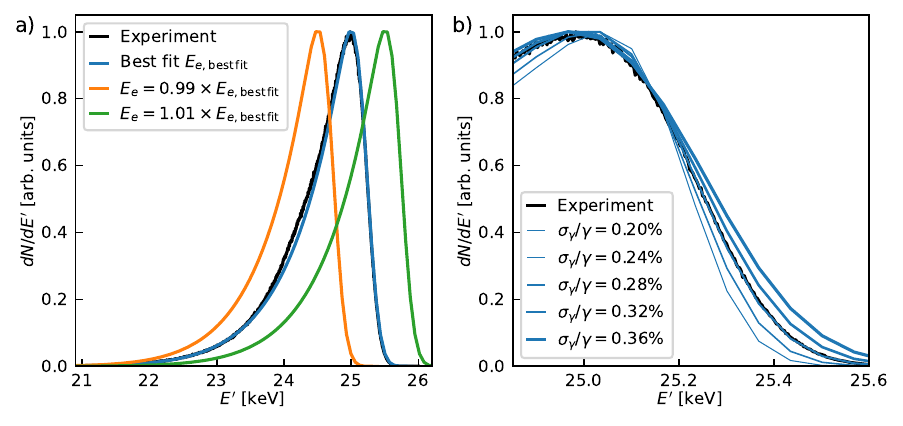}
\caption{Effect of the mean electron energy and electron beam energy spread on the X-ray spectrum at otherwise constant laser and electron beam parameters. a) Shift of the X-ray spectrum with variations of the mean electron energy.
b) Alteration of the X-ray spectrum's high-energy fall-off due to different electron beam energy spreads.}
\label{fig:energy}
\end{figure}

\subsection{How different electron beam parameters affect the peak energy and the spectrum's high-energy fall-off}  \label{sec_shape_spectra}
The X-ray spectrum's peak energy mainly depends on the mean electron beam energy for a given laser photon energy and a Gaussian electron energy distribution. Figure \ref{fig:energy} a) depicts the shift of the peak X-ray energy with changes in the mean electron energy for otherwise constant laser and electron parameters.
Since a \qty{\sim30}{\pico\second}-long (FWHM) Nd:YAG laser pulse is basically monochromatic, its energy spread, and in turn its contribution to a symmetric broadening of the X-ray spectrum, can be neglected.
In contrast, the electron energy spread is non-negligible.
For collimated colliding beams observed on-axis, the electron energy spread introduces a symmetric broadening around the peak X-ray energy.
If laser and electron beam are focussed, the electron beam emittance introduces a fall-off length on the low-energy side of the X-ray spectrum, which can slightly shift the peak of the X-ray spectrum and alter the high-energy fall-off.
Similarly, changing the electron beam's energy spread for the case of a non-negligible emittance affects slightly the peak X-ray energy in addition to the high-energy fall-off.
This effect is displayed in Figure \ref{fig:energy} b) and discussed in Section \ref{sec:x-ray} for otherwise constant laser and electron parameters.
Accordingly, the simultaneous presence of an electron energy distribution and emittance couples shifts of the peak position and alterations to the high-energy tail.
For this reason, the emittances, the energy spread and the mean energy of the electron beam need to be optimised simultaneously.
To this end, the genetic algorithm described in section \ref{sec:GA} was chosen for accomplishing this.

\subsection{Retrieving the electron beam parameters}
\begin{figure*}[]
\centering\includegraphics[width=0.9\textwidth]{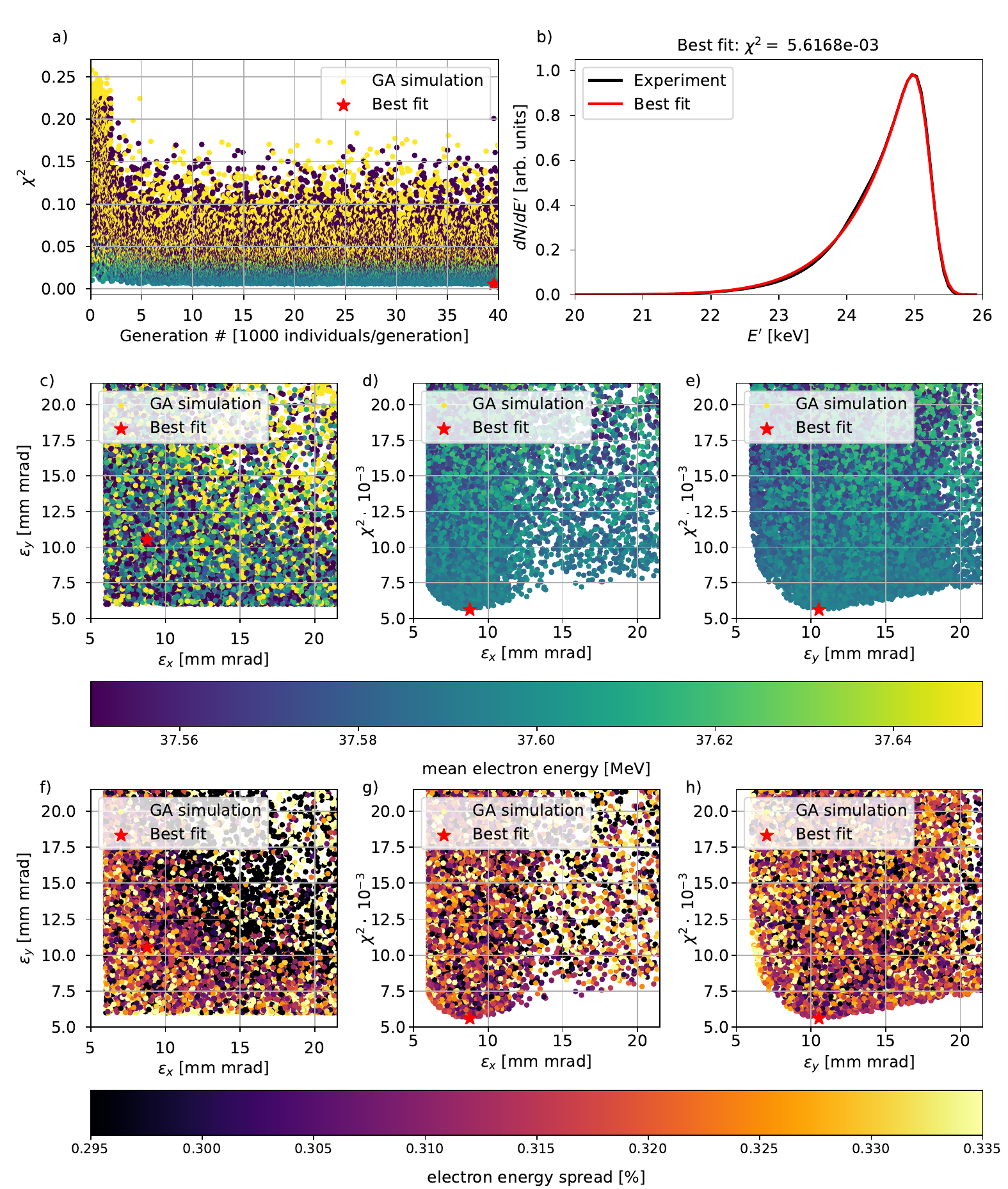}
\caption{Initial characterisation (iteration 1) of the MuCLS electron beam operated at the \qty{25}{keV} X-ray configuration by matching the calculated X-ray spectrum to the measured one.
a) shows the $\chi^2$-error for each individual of the genetic algorithm varying the horizontal and vertical electron beam emittances as well as the mean electron energy and the electron energy spread.
The mean electron energy is colour-coded.
The colourbar ranges from \qtyrange{37.55}{37.65}{\MeV}, i.e.\ it is centred around the best mean energy of the allowed range from \qtyrange{37.20}{38.20}{\MeV}.
The red star denotes the optimal fit found by the genetic algorithm after 40 generations with 1000 individuals. 
Its spectrum is plotted with the experimental data for comparison in b).
The emittance combinations employed by the genetic algorithm are plotted in c).
The $\chi^2$-error of the 2D search space projected onto the $\epsilon_x$-axis and the $\epsilon_y$-axis  is depicted in d) and in e), respectively.
The electron energy spread is colour-coded in f)-h), otherwise they are the same as c)-e).
The emittance combination resulting in the best fit in the initial characterisation is $\epsilon_x=\qty{8.8}{\milli\meter\milli\radian}$ and $\epsilon_y=\qty{10.5}{\milli\meter\milli\radian}$.
The emittance combination resulting in the best fit during all iterations is $\epsilon_x=\qty{9.0}{\milli\meter\milli\radian}$ and $\epsilon_y=\qty{10.8}{\milli\meter\milli\radian}$.
}
\label{fig:sp1}
\end{figure*}
\begin{figure*}[t!]
\centering\includegraphics[width=\textwidth]{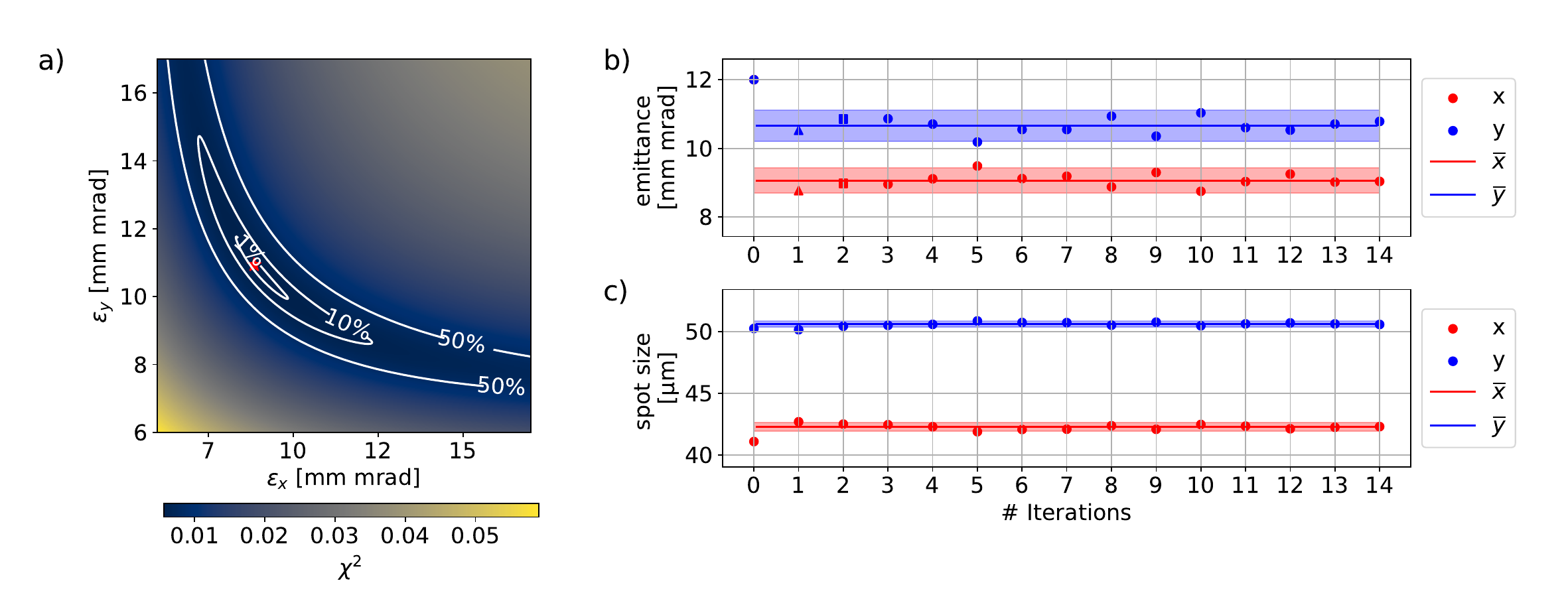}
\caption{
a) $\chi^2$-error map for a systematic electron beam emittance variation around the electron and laser parameter combination producing the best fit to the measured \qty{25}{\keV} spectrum in the iterative genetic algorithm.
The comparison was performed on a grid of 501 \texttimes\, 501 points.
The emittance combination that reproduces the X-ray spectrum best is indicated by a red star.
The white contours indicate areas in which the $\chi^2$ value increases by less than 1\%, 10\%, and 50\% compared to its minimum.
b) \& c) analyse the performance of the electron beam emittance and source size retrieval with respect to the number of iterations. 
b) depicts the emittance combinations resulting in the best match to the measured spectrum after each iteration of the iterative genetic algorithm while c) the corresponding source sizes for each iteration.
The coloured areas indicate the range within $\pm 2 \sigma$ of the mean values determined for iterations 2 to 14 depicted in panel b) \& c), respectively.
$\overline{x}$ and $\overline{y}$ are the corresponding  mean values.}
\label{fig:iterative_result}
\end{figure*}
The electron beam parameters are retrieved following the strategy outlined in Section \ref{sec:characterisation_strategy}, Point \ref{item:emittance_det}) - \ref{item:iteration_end}).
The electron beam's mean energy and its energy spread are allowed to vary in addition to electron beam emittances during this optimization, cf.\ Table \ref{tab:pulses}.
One of the cycles \ref{item:emittance_det}) - \ref{item:ebem_focus_update}) is called one iteration.
Over the first three iterations the boundaries for the emittance fit are reduced to sample the region closer to the minimum with a higher density of sampling points.
For further iterations, the boundaries are constant to avoid over-constraining the search-space in the genetic algorithm.
The number of individuals per generation is set to 1000 and the number of generations is 40.
Figure \ref{fig:sp1} depicts the first iteration of the emittance analysis with the genetic algorithm and the {\tt STARS} simulation code. The MuCLS was operated at the standard configuration producing an X-ray peak energy of about \qty{25}{\keV}.
The corresponding MuCLS parameters are listed in Table \ref{tab:pulses}.
In panel a), c) - h), each point represents one {\tt STARS} simulation run.
In Figure \ref{fig:sp1} a), c) - e), the colour encodes the mean electron energy restricted to the energy range between \qtyrange[range-phrase= { and }]{37.55}{37.65}{\MeV}.
In Figure \ref{fig:sp1} f) - h) the colour encodes the electron energy spread restricted to the range between \qtyrange[range-phrase= { and }]{0.295}{0.335}{\percent}.
The $\chi^2$-error is depicted for each simulation run for 40 generations of the genetic algorithm.
The simulation run with the lowest error is indicated with a red star.
Its spectrum already agrees well with the measurement, cf.\ Figure \ref{fig:sp1} b).
Figure \ref{fig:sp1} a) demonstrates that low $\chi^2$-values below $\approx 0.03$ can only be achieved when the mean electron energy is approximately correct, i.e.\ around \qty{37.6}{\MeV}.
Figure \ref{fig:sp1} c) displays the combinations of emittances and mean energy chosen in the genetic algorithm.
Figure \ref{fig:sp1} d) and e) show the projections of the $\chi^2$-value on both the horizontal and vertical emittance.
Although these graphs suggest the existence of a minimum at the position of the red star, this minimum is only reached if the electron beam energy spread is optimised as well.
This becomes obvious looking at Figure \ref{fig:sp1} g) and h).
After the first iteration, the retrieved mean electron energy is \qty{37.59}{\MeV}, electron energy spread is \qty{0.31}{\percent}, the horizontal emittance is \qty{8.8}{\mm \milli \radian} and the vertical emittance is \qty{10.5}{\mm \milli \radian}.
These emittances are now used to update the electron beam focus, cf.\ step \ref{item:ebem_focus_update}.
Afterwards, another iteration of the genetic algorithm starts.
For the-proof-of-principle 14 iterations were performed.
The optimal electron beam parameters retrieved after the iterative procedure are a mean electron energy of \qty{37.59}{\MeV}, electron energy spread of \qty{0.31}{\percent}, horizontal electron beam size of \qty{42.3}{\um} and vertical electron beam size of \qty{50.6}{\um}. The horizontal emittance is \qty{9.0}{\mm \milli \radian} and the vertical one is \qty{10.8}{\mm \milli \radian}.
Accordingly, mainly the emittances are slightly adjusted over the course of the subsequent iterations.
These parameters are also the ones that are most susceptible to changes in day-to-day operation of the MuCLS as they are most sensitive to the injection into the storage ring.\\
For an optimised mean electron energy and  electron energy spread, the projections of the $\chi^2$-value along the emittance axes, i.e.\ Figure  \ref{fig:sp1} g) \& h), indicate a single shallow valley for the optimal emittance combination.
\begin{figure}
    \centering
    \includegraphics[width=\linewidth]{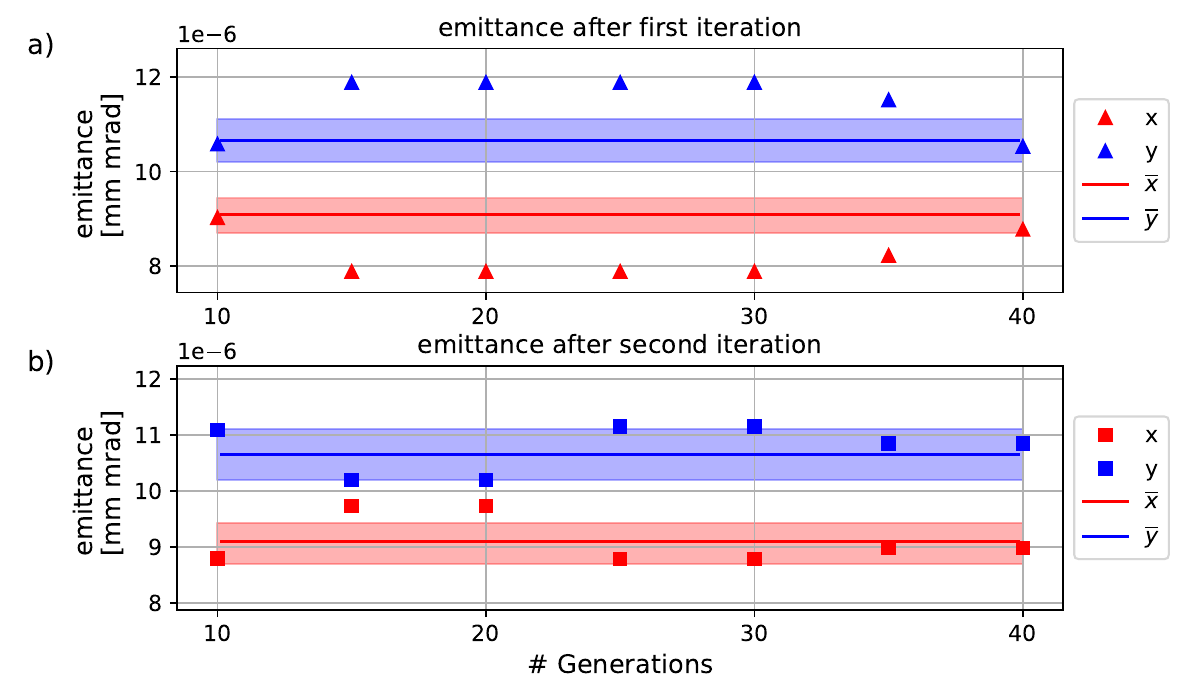}
    \caption{Dependence of the genetic algorithm on the number of generations.
    In a), the accuracy of the emittances returned after the first iteration of the genetic algorithm is analysed, while the emittances retrieved after the second iteration of the genetic algorithm are displayed in b).
    The coloured bands in a) and b) represent the same confidence interval depicted in panel \ref{fig:iterative_result} b) and $\overline{x}$ and $\overline{y}$ are the corresponding  mean emittances for iterations 2 to 14.
    If at least two iterations with 35 generations each are employed, the iterative genetic algorithm converges into the confidence interval.}
    \label{fig:dependence_on_generations}
\end{figure}
To evaluate this hypothesis, the emittance space was mapped systematically in the range between \qtyrange[range-phrase={ and }]{6}{17}{\mm\milli\radian} in both dimensions with a grid of 501~\texttimes~501 points.
Figure \ref{fig:iterative_result} a) is the resulting error map for the two-dimensional emittance space of the electron beam configuration retrieved above.
A hyperbolically shaped minimum can be recognised, supporting the hypothesis of a single minimum.
Both in $x$ and $y$ direction, the $\chi^2$-value increases steeply towards lower emittances, whereas it is much shallower towards higher emittances.
The contour lines mark the region in which the $\chi^2$-value increases by less than  1\%, 10\%, and 50\%, respectively, compared to the one of the optimal configuration indicated with a red star.
The minimum's diagonal and almost symmetric orientation can be explained by the fact that slight symmetric changes of the emittances have the same effect on the spectrum. 
Figure \ref{fig:iterative_result} b) displays the optimal parameters retrieved after each iteration for a run with 40 generations (1000 individuals each) per iteration.
The retrieved emittances remain within the confidence interval directly after iteration 1.
Figure \ref{fig:iterative_result} c) depicts the electron beam source size for each iteration.
In general, the source size changes significantly only after the fist iteration.
As soon as two iterations are performed, the source sizes converge into the narrow confidence interval.
The confidence interval in b) and c) is defined as $ \bar\epsilon_{x/y}\pm 2\sigma$ or $\bar\sigma_{e,x/y}\pm 2\sigma$, respectively.
The mean emittance $\bar\epsilon_{x/y}$ and the mean electron spot size $\bar\sigma_{e,x/y}$ are calculated for iterations 2-14.\newline
Following up on this result, the required number of generations per iteration was investigated.
Figure \ref{fig:dependence_on_generations} a) \& b) display the retrieved emittances after iteration number one and after iteration number two, respectively.
If one iteration is run, the emittances converge only to the confidence interval when 40 generations are employed.
In contrast, the emittance converges to the confidence interval if more than 25 generations are run for two iterations.
Nevertheless, 35 generations improve the result further, cf.\ Figure \ref{fig:dependence_on_generations} b).
Three iterations with 25 generations improve the retrieved emittances similarly.
These results allow sampling the search space much more coarsely in the future, especially for the mean electron beam energy and its energy spread, speeding up the analysis while still ensuring convergence of the emittances.

\begin{table}[h]
    \caption{Summary of relevant MuCLS electron, laser and aperture parameters for this study.
    Quantities for beam sizes, pulse durations and energy spread are rms-values.\newline
    $^{*}$ parameters allowed to vary during the iterative genetic algorithm}
    \begin{tabular}{llll}
        \hline\hline 
        Quantity & Unit & Value  \\
        \hline
        X-ray horizontal spot size $\sigma_{\mathrm{X-ray},x}$ & $\si{\um}$  & 42.6 \\
        X-ray vertical spot size $\sigma_{\mathrm{X-ray},y}$ & $\si{\um}$  & 46.3 \\
        X-ray flux  & $\si{ph/s}$  & $0.73 \cdot 10^{10}$ \\
        \hline
        Electron beam energy $E_\gamma$ & MeV & $37.59$ \\
                                        &     & $[37.2, 38.2]^{*}$\\
        Electron beam energy spread $\sigma_{E_\gamma}/E_\gamma$     &                 & $0.0031$ \\
                                                               &                 &$[0.002, 0.004]^{*}$ \\
        Electron beam horizontal spot size $\sigma_{\gamma,x}$ & $\si{\um}$      & $42.3^{*}$ \\
        Electron beam vertical spot size $\sigma_{\gamma,y}$   & $\si{\um}$      & $50.6^{*}$  \\
        Normalized horizontal emittance $\epsilon_x$           & ${\rm m~rad}$ & $9.0\cdot 10^{-6}$ \\
        & & $[6, 25]\cdot 10^{-6*}$ \\
        Normalized vertical emittance $\epsilon_y$             & ${\rm m~rad}$ & $10.8\cdot 10^{-6}$ \\
        & & $[6, 25]\cdot 10^{-6*}$ \\
        Electron beam pulse duration $\tau_{e}$                & ${\rm ps}$      & $50$ \\
        Electron beam pulse charge & ${\rm pC}$ & $240$ \\
        \hline
        Laser wavelength $\lambda$                             & $\si{\um}$      & $1.064$ \\
        Laser horizontal spot size $\sigma_{l,x}$              & $\si{\um}$      & $65.4$  \\
        Laser vertical spot size $\sigma_{l,y}$                & $\si{\um}$      & $89.9$ \\
        Laser pulse duration $\tau_{l}$                      & ${\rm ps}$      & $11.2$ \\
        Normalized length $s\equiv c\tau_l/\lambda$         &                 & $3155$ \\
        Laser energy in cavity                                 &$\si{\kilo\watt}$& $104.4$ \\
        \hline
        Radius of the circular aperture for\\ spectrum measurement $r_a$                  & ${\rm mm}$      & $2.525$ \\     
        Distance $L$ between IP and aperture $r_a$                                     & ${\rm m}$       & $5.2$\\
       \hline\hline
    \end{tabular} \label{tab:pulses}
\end{table}


\subsection{X-ray flux estimation}
Once all the laser and electron beam parameters at their interaction point have been determined, the expected X-ray flux can be calculated based on them.
The {\tt STARS} code does not include flux reduction due to the overlap of extended beams, the so-called hourglass effect.
Accordingly, this effect is assessed for the case of the MuCLS to account for its flux reduction.

\subsubsection{Reduction in brilliance due to the hourglass effect}
The three-dimensional focus geometry results in a geometric reduction of the scattering rates, called the hourglass effect \cite{furman1991}.
It is accounted for by computing the brilliance reduction factor $R_{\rm hg}$, as outlined by Furman et al.\ \cite{furman1991}, for the MuCLS parameters.
The reduction is less than \SI{6.4}{\percent}.

\subsubsection{Expected X-ray flux}
The X-ray flux is measured behind the MuCLS's exit window while {\tt STARS} simulates MuCLS's flux at the interaction point by integrating the spectro-angular flux across the opening angle of the aperture and the total spectral range.
Since the opening angle is asymmetric, cf.\ \cite{Eggl2016}, a circular aperture with the mean opening angle of both directions was employed in the calculation to approximate the the flux transmitted through the real aperture.
In between the X-ray source spot and the measurement position, the back-thinned laser mirror forming the angular aperture and the beryllium vacuum window of the laser resonator are located.
Multiplying the spectro-angular flux density with their energy-dependent absorption function before the aforementioned integration transfers the flux calculated at the interaction point to the one at the location of the measurement.
With a nominal charge of \SI{240}{\pico \coulomb} inside the storage ring and a mean stored power of \SI{\sim105}{\kW}, {\tt STARS} suggests a flux of \SI{2.27e10}{\photon \per \s} at this position.
Incorporation of the hourglass effect reduces the calculated flux to \SI{2.13e10}{\photon \per \s}.

\section{Potential realisation of an electron beam monitor and discussion of current limitations}
In this study, a framework for the characterisation of an ICS is presented.
One key innovation achieved in this context is the establishment of a physics framework for the computation of Compton scattering in the linear regime, called {\tt STARS}, that reduces computation times by orders of magnitude compared to the previous code {\tt ICCS3D} described in \cite{ranjan18}.
This achievement allows for employing a genetic algorithm for the retrieval of several electron beam parameters, like its emittances, mean energy, and energy spread, from the measurement of an ICS's spectrum and the properties of the colliding laser.
If the invasive ringdown technique is replaced with a passive, non-invasive monitoring as suggested by Locke et al.\ \cite{Locke2009}, the proposed framework transforms into a fully non-invasive one if the spectrum is measured non-invasive.
As a consequence, a tool for electron beam quality monitoring would become available.\\
For real-time monitoring, the time for the analysis must be reduced to the second scale.
Since systematic sampling of the parameter space revealed only one global minimum, the genetic algorithm could be replaced by a classical gradient-based minimisation, like steepest descent.
Hence, real-time emittance monitoring will become feasible using such solvers.
This is very useful for the operators during tuneup of the electron beam and may even be employed in automated feedbacks.\\
One limitation in the current stage of the proposed characterisation technique is that the X-ray flux cannot be employed as a fitting parameter.
The simulated flux of \qty{2.13e10}{\photon \per \s} is about a factor of $2.9$ larger than the measured one of \qty{0.73e10}{\photon \per \s}.
However, some effects may introduce an error that has not been considered so far.
First, the calibration of the charge monitors at the MuCLS may no longer be perfect as they have not been recalibrated since the machine installation.
If the MuCLS is operated at a nominal charge of \qty{240}{\pico\coulomb}, the different cavity monitors display a charge inside the ring of \qtyrange{210}{255}{\pico\coulomb}.
This adds an uncertainty of at least between \qtyrange[range-phrase={ and }]{-12.5}{6.25}{\percent}.
In addition, the electron bunch length could deviate from the specified value, and timing jitters between the laser and the electron beam could significantly affect the overlap of both beams.
Unfortunately, neither assessing the timing jitter nor determining the electron bunch length is currently possible at the MuCLS because of a lack of diagnostic capabilities.
On the laser side, the laser power stored inside the cavity is calculated based on information retrieved by built-in diagnostics, which have not been recalibrated since Lyncean Technologies Inc.\ went out of business.
Further, the transmissivity of the enhancement cavity's input coupling mirror was provided by Lyncean Technologies Inc.\ as well.
This increases the uncertainty on the measurement side and needs to be addressed for properly validating the X-ray flux.
To this end, the diagnostics at the MuCLS are going to be re-calibrated, enhanced, and missing diagnostic capabilities for the laser and electron beam are going to be developed.
Furthermore, {\tt STARS} does not yet include a non-zero crossing angle or an extended interaction region which are going to be included as a next step.
If subtle differences between the X-ray flux predictions by {\tt STARS} and the  measurements remain once this has been achieved, understanding their physical origin will be important to optimise and improve ICSs further.
To generalise the model further, a rotation of the emittance ellipse is going to be implemented which is required to model extended beams.\\
So far, the discussion was restricted to ICSs.
Nevertheless, the characterisation of the electron beam via measurement of the (inverse) Compton-scattered X-ray spectrum can be useful for all types of accelerators, e.g.\ as a non-invasive emittance monitor inside the storage ring of a synchrotron.
In this case, a simple low-power monochromatic continuous-wave laser can be employed because the X-ray flux required for a diagnostic tool is many orders of magnitude lower than for imaging experiments.
In summary, this study presented a very powerful framework for the characterisation of an inverse Compton X-ray source with a focus on the retrieval of the electron beam parameters.
To this end, a physical model of the ICS process was established, allowing the rapid calculation of X-ray spectra from (inverse) Compton-scattered laser photons.
The electron beam characterisation can be employed at any type of accelerator and may be used as a tool for electron beam monitoring also elsewhere, e.g.\ at synchrotrons.

\section*{Funding}
This work was supported by the Centre for Advanced Laser Applications (CALA) and the Deutsche Forschungsgemeinschaft (DFG, German Research Foundation, grant number: $513827659$).
B.~T.~acknowledges the support from the U.S.~National Science Foundation CAREER award No.~1847771 and from the U.S.~National Science Foundation CSSI award No.~2513760.
This work is authored by Jefferson Science Associates, LLC under U.~S.~Department of Energy (DOE) Contract No.~DE-AC05-06OR23177.

\section*{Acknowledgments}
This work is authored by Jefferson Science Associates, LLC under U.~S.~Department of Energy (DOE) Contract No.~DE-AC05-06OR23177.
The U.~S.~Government retains a non-exclusive, paid-up, irrevocable, world-wide license to publish or reproduce this manuscript for U.~S.~Government purposes.

\section*{Disclosures}

\noindent The authors declare no conflicts of interest.

\section*{Data availability statement}
\noindent The data underlying the results presented in this paper is not publicly available at this time but may be obtained from the authors upon reasonable request.\\

\appendix
\section{Comparison of STAR to ICCS3D and CAIN\label{app:comp_STARS}}
\begin{figure} [H]
    \centering
    \includegraphics[width=\linewidth]{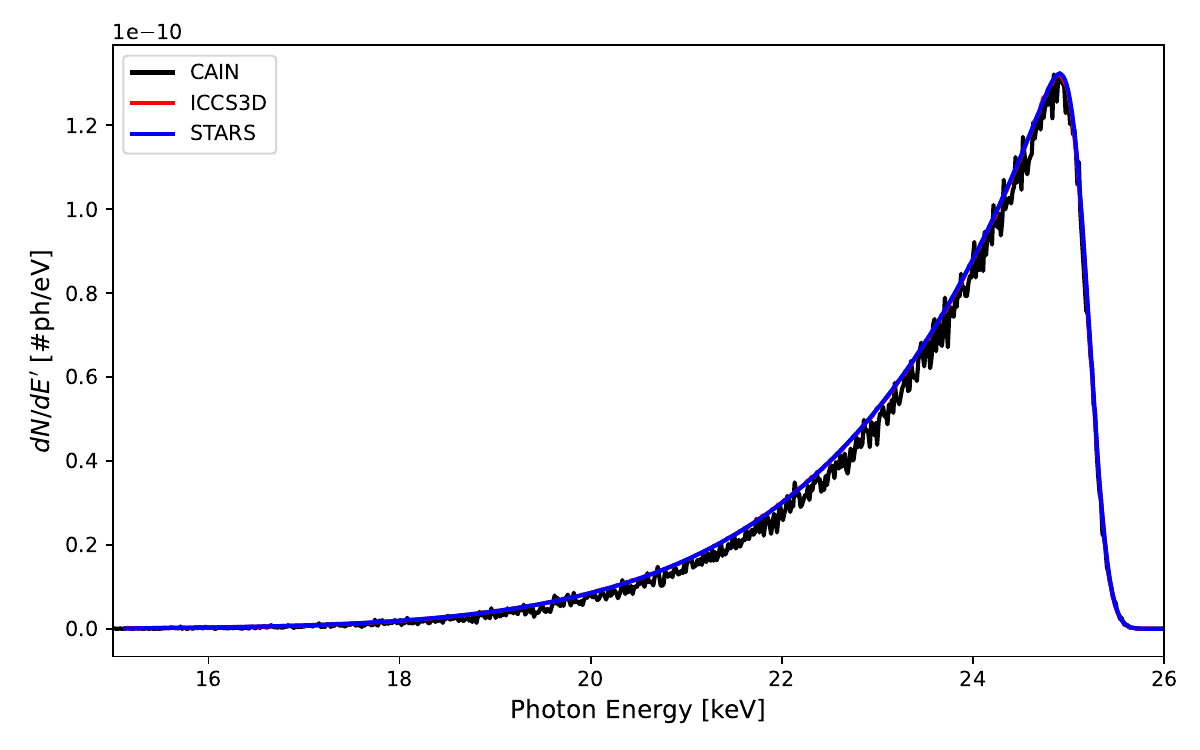}
    \caption{Comparison of the X-ray spectra calculated with STARS, ICCS3D and CAIN. All spectra agree very well.}
    \label{fig:comparison_spectra}
\end{figure}

The main manuscript introduces a new approach to efficiently calculate radiation of Compton scattering X-ray sources operating in the linear regime.
Naturally, the Compton sources radiation properties must be independent of the used mathematical formulation or code, respectively.
To this end, the X-ray spectrum and the expected X-ray flux are calculated for the same electron and laser beam parameters, cf.\ Table \ref{tab:pulses} in the main manuscript, with our new code STARS, ICCS3D \cite{krafft16,ranjan18} and CAIN \cite{CAIN}.
Figure \ref{fig:comparison_spectra} displays the X-ray spectra calculated with the different codes for the MuCLS aperture.
All spectra agree perfectly, apart from small numerical noise typical for CAIN (see discussion in Section III.B in Ref.~\cite{ranjan18}).
STARS retrieves an X-ray flux of \SI{2.27e10}{ph/\s}, ICCS3D one of \SI{2.27e10}{ph/\s} and CAIN one of \SI{2.27e10}{ph/\s}.
Only very minor differences exist.

\section{Influence of electron beam emittance and focus size on the X-ray spectrum\label{app:influence_emittance_focus}}
\begin{figure}[h]
\centering\includegraphics[width=\linewidth]{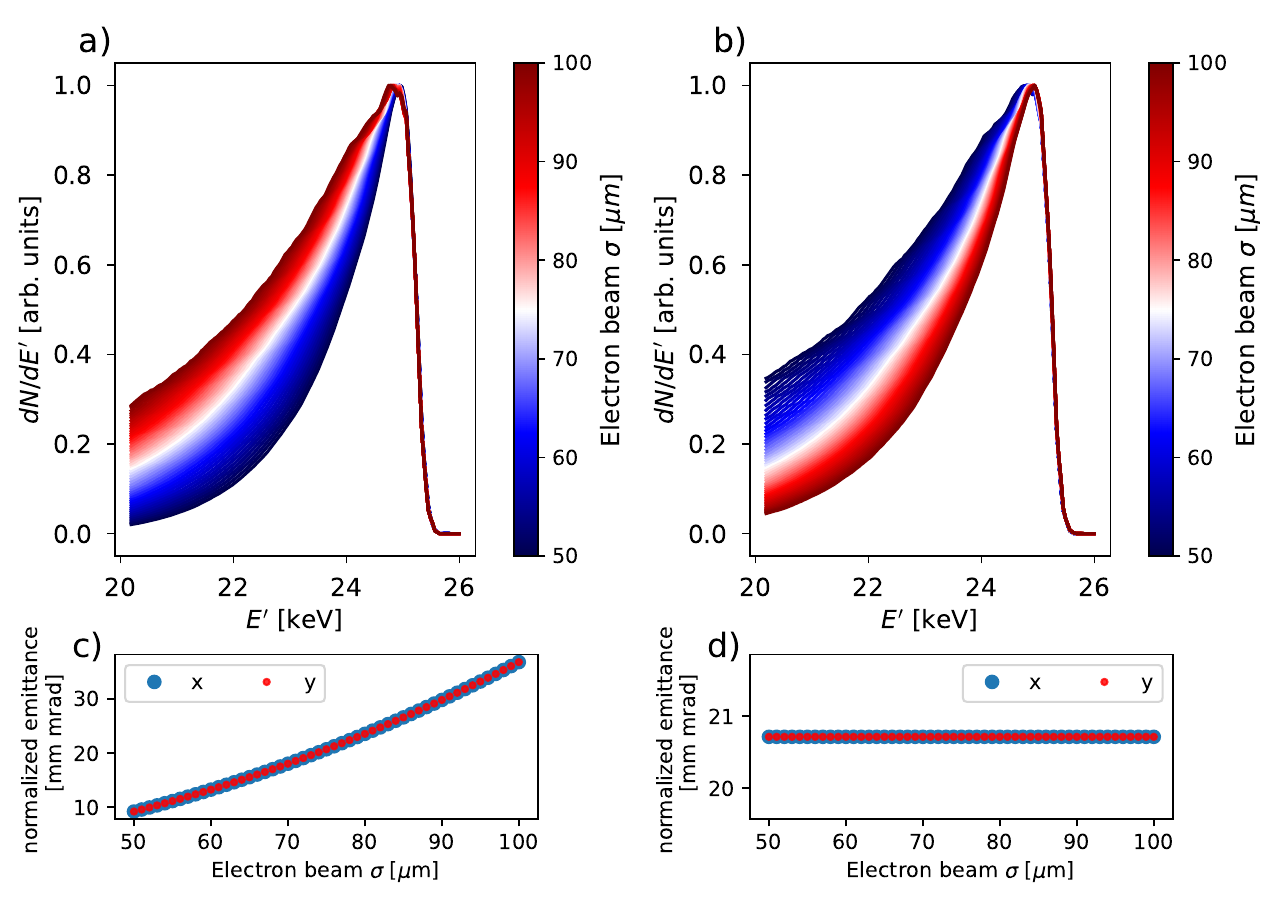}
\caption{Dependence of the X-ray spectrum at the interaction point on changes of the electron beam focus for different cases.
In a), the spectra are calculated with the electron beam focal sizes and emittances plotted in c), i.e., with a constant beta function of $\beta=\qty{2.0}{\cm}$.
In b), the spectra are depicted for increasing electron beam size while keeping the symmetric emittance of \qty{20.7}{\mm \milli\radian} constant, cf.\ d).
In both cases, the laser parameters are fixed to a round focus size with \qty{50}{\um} (rms).
Furthermore, both, emittances and X-ray spectra are colour-coded with the electron beam size used for the specific simulation.
}
\label{fig:supp_spectrum_emittance_fixed}
\end{figure}

Figure \ref{fig:supp_spectrum_emittance_fixed} displays the influence of changes of the electron beam emittance and electron focus size on the X-ray spectra for constant laser parameters and a fixed aperture for the X-ray beam.
The observed alteration of the spectrum does not originate from changes of the X-ray source size which increase the effective X-ray opening angle of a fixed aperture only by a few micro-radians.
Instead, changes in the electron beam size affect the distribution of the transverse momentum of the electrons at the interaction point if either the beta function or the emittance is kept constant.
Accordingly, the X-ray spectrum emitted into the same X-ray cone angle is altered.

Figure \ref{fig:supp_spectrum_emittance_fixed} demonstrates this effect for two different situations:
\begin{itemize}
    \item The electron beam’s beta function is kept constant and the electron beam size is varied.
    Accordingly, the electron beam emittance increases with increasing electron beam size.
    As a result, the transverse momentum spread of the electron beam increases at the interaction region.
    The higher the emittance of the electron beam, the broader the low-energy fall-off of the X-ray spectrum.
    This is depicted in Figure \ref{fig:supp_spectrum_emittance_fixed}a).
    The corresponding emittances for each spectrum are presented in Figure \ref{fig:supp_spectrum_emittance_fixed} c).
    Both, emittances and X-ray spectra are colour-coded with the electron beam size used for the specific simulation.
    \label{item:i}
    \item The electron beam’s emittance is kept constant and the electron beam size is varied.
    Hence, the electron beam’s divergence decreases with increasing electron beam size, or, in other words, the electron beam’s beta function increases.
    Contrary to situation \ref{item:i}, the electron beam’s transverse momentum spread decreases in the interaction region.
    Consequently, the X-ray spectrum's low-energy fall-off narrows with increasing electron beam size at a constant emittance.
    This is depicted in Figure \ref{fig:supp_spectrum_emittance_fixed} b).
    The corresponding emittances for each spectrum are presented in Figure \ref{fig:supp_spectrum_emittance_fixed} d).
\end{itemize}
Since both, changes of the electron beam size at a constant emittance and changes of the electron beam size at a constant beta function, affect the emitted X-ray spectrum, iterating between the genetic algorithm updating the electron beam emittance, step \ref{item:emittance_det}, and the electron beam size determination based on the projected X-ray source size, step \ref{item:ebem_focus_update}, is necessary.

\section{Estimation of the interaction angle\label{app:interaction_angle}}
The centres of the mirrors M1 and M4, cf.\ Figure 2 a), are aligned with the electron beam axis in the straight section of the storage ring.
Hence, the electron beam axis and the optical axis of the enhancement cavity are aligned in the interaction region.
Consequently, the back-thinned area in the centre of mirror M4 through which the X-rays are transmitted is centred well on the electron beam axis.
However, the laser beam cannot be placed on the back-thinned area. 
Therefore, it is shifted sideways to hit right next to this area to keep the interaction angle as small as possible.
\begin{figure}
    \centering
    \includegraphics[width=0.75\linewidth]{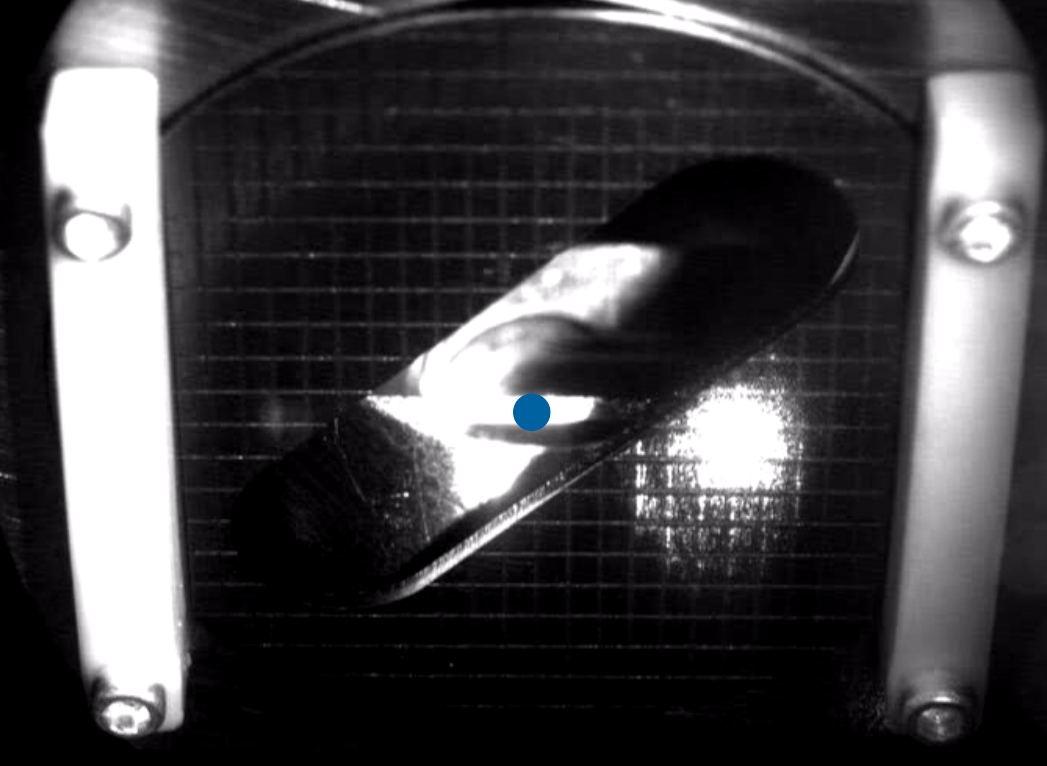}
    \caption{Insert of the enhancement cavity located \qty{1}{\m} in front of the interaction point, displaying the laser beam offset from the centre of the optical axis. The displacement is about \qty{6}{\mm}, resulting in an interaction angle of about \qty{6}{\milli\radian}.}
    \label{fig:beam_on_insert}
\end{figure}
\FloatBarrier
The interaction angle was estimated in two ways:
\begin{enumerate}[nosep]
    \item The back-thinned area of the mirror forms the X-ray aperture.
    The full divergence angle of the X-ray beam is \qty{\sim 4.5}{\milli \radian}. 
    Hence, the half-angle is \qty{\sim 2.25}{\milli \radian}. 
    Mirror M4 is located at a distance of \qty{\sim 1.25}{\m} from the interaction point.
    Accordingly, the radius of the back-thinned area is about \qty{2.9}{\mm}. 
    The laser waist on M4 is \qty{\sim 3}{\mm} as well.
    Accordingly, the laser beam center should be located twice the waist away from the back-thinned area, i.e., \qty{\sim 9}{\mm}. 
    Taking into account the distance of the mirror from the interaction point, the interaction angle would be around \qty{7}{\milli \radian}, or \qty{0.4}{\degree}.
    \item A screen with a mm-grid exists to monitor the laser beam position after M4.
    It is positioned at a distance of \qty{1}{m} from the interaction point.
    The centre of the laser beam is at about \qty{6}{\mm} from the centre, cf.\ Figure \ref{fig:beam_on_insert}.
    This would result in a collision angle of about \qty{6}{\milli \radian}, or \qty{0.34}{\degree}.
\end{enumerate}

In conclusion, the interaction angle is about \qtyrange{6}{7}{\milli \radian}.
Since, the position measurement is more reliable compared to the heuristic estimation, an angle of \qty{6}{\milli \radian} was used in the retrieval of the electron beam size from the measured projected X-ray source size, step \ref{item:ebeam_focus}.

\section{Alignment of the spectrometer\label{app:alignment_spectrometer}}
\begin{figure}
    \centering
    \includegraphics[width=\linewidth]{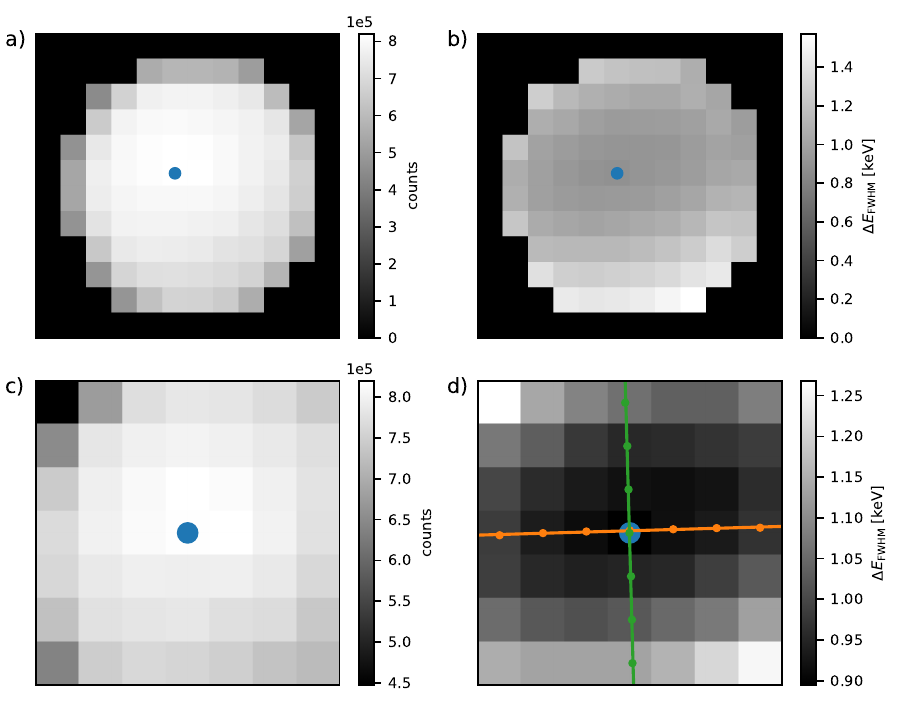}
    \caption{Alignment of the spectrometer. a) displays the integrated  intensity, b) the full-width at half maximum (FWHM) bandwidth of the X-ray beam. c) and d) are the respective zooms onto the centre of the radiation cone, indicated as a blue circle. It was determined as the position of highest intensity at smallest bandwidth. This is the position used for the spectrum measurements.}
    \label{fig:int_fwhm_spectrum}
\end{figure}

The detector was aligned with the centre of the X-ray beam by raster-scanning the spectrometer in 2D with a step size of \qty{2.0}{\mm} over an area larger than the X-ray beam.
The spectrum was integrated at each position.
The resulting energy-integrated intensity is displayed in Figure \ref{fig:int_fwhm_spectrum} a).
Additionally, the spectrum's full-width at half-maximum (FWHM) was calculated.
The FWHM is displayed in Figure \ref{fig:int_fwhm_spectrum} b) for the full beam.
The position of highest spectrally integrated intensity at the smallest bandwidth was identified as the centre of the radiation cone and, in turn, used as the position of the spectrometer for the spectrum measurements.
c) shows a zoom-in of a) on the centre.\\
Two axes with slightly differently increasing bandwidth are visible in the zoom-in on the centre of the radiation cone, Figure \ref{fig:int_fwhm_spectrum} d).
If the electron beam emittance ellipse is rotated with respect to the coordinate system, the X-ray spectrum's bandwidth should follow this rotation.
In other words, by raster-scanning the X-ray beam, two axes should be identifiable if two different emittances exist, one along which the electron beam bandwidth is largest and one along which it is narrowest.
A stronger increase of the X-ray bandwidth can be observed along the vertical direction than along the horizontal one in Figure \ref{fig:int_fwhm_spectrum} d).
Furthermore, the axes with narrower and broader X-ray bandwidth appear to be rotated slightly with respect to the laboratory coordinate system.
To quantify the rotation angle of the ellipse's long axis, the centre-of-mass of the X-ray beam's FWHM bandwidth was calculated for each row, cf.\ green dots and the respective linear fit as a green line in Figure \ref{fig:int_fwhm_spectrum} d).
The position of the minimum was calculated by multiplying the weights (FWHM bandwidths) with $-1$ in the centre-of-mass calculation for each column, cf.\ orange dots and the respective linear fit in orange in Figure \ref{fig:int_fwhm_spectrum} d).
The rotation angle in both cases is about \qty{1.6}{\degree}.
This angle is considered to be sufficiently small to approximate the emittance ellipse to be aligned with the coordinate axes.

\section{Assumption of thermal effects on enhancement cavity mirrors\label{app:thermal_effects}}
In equilibrium, a constant power is stored in the enhancement cavity.
All mirrors have the same thickness, except for the small back-thinned area on mirror M4.
However, no laser beam is placed on this thinned area.
In this situation, the temperature difference between the inner surface of the mirror and its back side is proportional to 1/thermal conductivity.
On the other hand, the change in thickness of a mirror is proportional to the thermal expansion coefficient times the difference in temperature.
Accordingly, the expansion is proportional to the thermal expansion coefficient / thermal conductivity.
For the two mirror substrates ULE \cite{ULE} and single crystal silicon \cite{Glassbrenner1964,Middelmann2015}, these ratios are approximately the same (ULE $\propto \sim 2\cdot10^{-8}/$ 1.31 [1/K /(W/(m K))]; Si $\propto \sim 2.55\cdot10^{-6}/150$ [1/K /(W/(m K))]).
Consequently, the expansion behaviour of both materials is very similar.
Further, the beam size on all mirrors is comparable, cf.\ Figure 2 b).
Accordingly, the distribution of the thermal load on the mirrors is comparable.
In turn, assuming the same deformation of all the mirrors is a reasonable approximation.

\FloatBarrier

\bibliography{paper}

\end{document}